\documentclass[%
reprint,
superscriptaddress,
preprintnumbers,
 amsmath,amssymb,
 aps,
prl,
]{revtex4-2}

\usepackage{subfigure}
\usepackage{graphicx}% Include figure files
\usepackage{dcolumn}% Align table columns on decimal point
\usepackage{bm}% bold math
\usepackage{color}
\usepackage[normalem]{ulem}
\usepackage{hyperref}% add hypertext capabilities
\hypersetup{colorlinks=true,allcolors=blue}
\begin{document}
\title{Spin Transport in a Quantum Spin-Orbital Liquid}
\author{Zekun Zhuang}
\affiliation{Department of Physics,   Brown University, Providence, Rhode Island 02912-1843, USA}
\author{J. B. Marston}
\affiliation{Department of Physics,   Brown University, Providence, Rhode Island 02912-1843, USA}
\affiliation{Brown Theoretical Physics Center, Brown University, Providence, Rhode Island 02912-1843, USA}

\date{\today }

\begin{abstract}
Quantum spin-orbital liquids (QSOLs) are a novel phase of matter, similar to quantum spin liquids, with quantum fluctuations in both spin and orbital degrees of freedom. We use non-equilibrium Green's function theory to study out-of-equilibrium spin transport in an exactly solvable QSOL model put forward by Yao and Lee. We find that the spin transport problem can be mapped to that of a free fermion problem with effective fermionic baths that have rapidly varying density of states. In the gapless phase, the spin current $I_s-V_s$ relation is thus highly nonlinear, while in the chiral gapped phase, the spin current conductance is quantized to be $1/2\pi$ provided that the contacts are sufficiently wide. The quantized conductance is a signature of the topological nature of the chiral gapped QSOL.
\end{abstract}

%\pacs{}
\maketitle


%\label{sec:level1}First-level heading:\protect\\ The line
%break was forced \lowercase{via} \textbackslash\textbackslash}

\textit{Introduction.}---Quantum spin liquids (QSLs) are a form of matter with no long-range order, long-range entanglement, fractionalized excitations and emergent gauge fields \cite{Savary2016,Zhou2017,broholm2020}. Owing to an exactly solvable model of a QSL introduced by Kitaev and a seminal paper by Jackeli and Khaliullin \cite{Kitaev2006, Jackeli2009} which points out that the model can be realized in some strongly spin-orbit coupled materials, Kitaev-type QSLs are an active area of investigation. The Kitaev model has also been generalized to spin-orbital models, or Kugel-Khomskii models \cite{kugel1973,kugel1982}, which have both spin and orbital degrees of freedom on each site.  Yao and Lee derived a SU(2)-symmetric version of such a model which lacks spontaneous symmetry breaking in its ground state and has novel excitations, such as non-Abelian spinons and fermionic magnons (FMs) \cite{Yao2011}. Such novel quantum phases, termed quantum spin-orbital liquids (QSOLs), are proposed to exist in a broader range of Kugel-Khomskii models \cite{feiner1997,oles2000,Yao2009,Corboz2012,Nakai2012,nussinov2015,Natori2016,Natori2018,Natori2019,Natori2020}.  Candidate materials that may realize the QSOLs, such as $\text{Ba}_3\text{CuSb}_2\text{O}_9$ \cite{Zhou2011, Nakatsuji2012,Quilliam2012,ishiguro2013,Smerald2014,Katayama2015,Smerald2015}, are still under investigation, while additional candidates may be found in certain $4d^1$ or $5d^1$ Mott insulators and twisted superlattice systems \cite{Chen2010,Natori2016,Natori2018,Yamada2018,Ushakov2020,Venderbos2018,Natori2019}.

Experimental confirmation of QSLs and QSOLs has been a long-standing problem. Because QSLs are electrically insulating, well-developed transport techniques cannot be utilized except in a few cases \cite{Aasen2020,Konig2020}. Thermal transport experiments have been useful for identifying QSLs--especially topological QSLs, such as $\alpha-\text{RuCl}_3$ in a magnetic field, where half-integer quantized thermal Hall conductance has been reported \cite{kasahara2018}. Theory and experiment of spin transport in QSLs or QSOLs is less well developed \cite{Hirobe2017,Hirobe2018,Koga2020,Minakawa2020,Mizoguchi2020,han2020}. Chen \textit{et al.} \cite{Chen2013} and Chatterjee \textit{et al.} \cite{Chatterjee2015} suggested that by sandwiching a QSL material between two paramagnetic metals and driving a spin current through the structure, one could characterize different types of QSLs as they have different power laws of spin current $I_s-V_s$ relation. De Carvalho \textit{et al.} generalize the results to the Yao-Lee QSOL model, and their calculations show that in the gapless phase with zigzag-type contact $I_s\sim V_s$ while in the chiral gapped phase $I_s\sim V_s^3$ \cite{DeCarvalho2018}. All the prior spin transport calculations use equilibrium spin correlation functions.

In this Letter, we relax the assumption that the system is near equilibrium and give a systematic formulation of the spin transport problem in a clean Yao-Lee model using non-equilibrium Green's functions (NEGFs). 
The difficulty of treating spin operators in diagrammatic approaches due to their non-commutativity can be tackled using Majorana representation \cite{Spencer1967,Coleman1993,Shnirman2003, Mao2003}. 
We find that the original spin transport problem may be mapped to a free fermion transport problem in the presence of fermionic baths with different effective temperatures, chemical potentials, and non-constant density of states (DOS). For the gapless phase, the transport characteristic is highly nonlinear: for the zigzag-type contact (ZC) $I_s\sim V_s^3$ while for the armchair-type contact (AC) $I_s\sim V_s^5$. For the chiral gapped phase, the spin current conductance is quantized if the contact is wide enough.  Our results for the gapless phases with ZCs and chiral gapped phases differ from those found in Ref. \cite{DeCarvalho2018} because we account for non-equilibrium accumulation of spin excitations. Such spin transport experiments can detect the topological phases of QSOLs and test the existence of the predicted FMs.

\textit{Model.}---Yao and Lee constructed a SU(2)-symmetric spin-$1/2$ model on the decorated honeycomb lattice \cite{Yao2011}. Despite the complexity of the original Hamiltonian and its underlying lattice geometry, the low-energy physics is described by a Kitaev-type Hamiltonian on the honeycomb lattice as shown in Fig. \ref{fig:lattice}:\begin{figure}[t]
    \centering
    \includegraphics{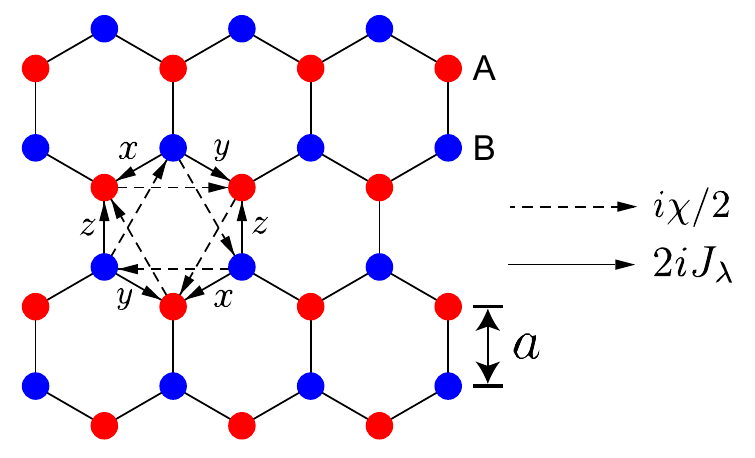}
    \caption{Yao-Lee model on the honeycomb lattice. The type $\lambda$ of bond is denoted by $x,y,z$ in the figure. An arrow from site $j$ to $i$ indicates matrix element $h_{ij}$ of the Hamiltonian of Eq. (\ref{magnonMajoranaH}) for $u_{ij}=1$.  A solid arrow indicates $h_{ij} = 2i J_\lambda$ while a dashed line indicates $h_{ij} = i\chi/2$.}
    \label{fig:lattice}
\end{figure}
\begin{equation}
    \begin{split} H_{\text{tot}}&=
\sum_{\langle ij\rangle_\lambda}\frac{J_\lambda}{4}  [\tau_i^\lambda\tau_j^\lambda][\vec{\sigma_i}\cdot\vec{\sigma_j}]\\&+\frac{\chi}{4}\sum_{\langle ij\rangle_\alpha\langle jk\rangle_\beta}\epsilon^{\alpha\beta\gamma}[\tau_i^\alpha\tau_j^\gamma\tau_k^\beta][\vec{\sigma_i}\cdot\vec{\sigma_k}].   
    \end{split} \label{OriginalH}    
\end{equation}
Pauli matrices $\vec{\sigma_i}=(\sigma^x_i,\sigma^y_i,\sigma^z_i)$ describe the spin degrees of freedom on site $i$, while $\tau^{x,y,z}_i$ are for the `orbital' degrees of freedom. The nearest-neighbor bond between site $i$ and $j$ of type $\lambda=x,y,z$ is denoted by $\langle ij\rangle_\lambda$ and particularly $\langle ij\rangle_\alpha\langle jk\rangle_\beta$ labels three neighboring sites $i,j,k$ that are ordered clockwisely within the corresponding plaquette. This model can be exactly solved by representing Pauli matrices in terms of Majorana fermions $\sigma_i^\alpha=-\frac{\epsilon^{\alpha \beta \gamma}}{2}i\gamma_i^\beta \gamma_i^\gamma$, $\tau_i^\alpha=-\frac{\epsilon^{\alpha \beta \gamma}}{2}id_i^\beta d_i^\gamma$, where $\gamma^{x,y,z}$ and $d^{x,y,z}$ satisfy anticommutation relations $\{\gamma_i^\alpha,\gamma_j^\beta\}=2\delta_{ij}\delta^{\alpha\beta}$, $\{d_i^\alpha,d_j^\beta\}=2\delta_{ij}\delta^{\alpha\beta}$ and $\{\gamma_i^\alpha,d_j^\beta\}=0$ \cite{Kitaev2006,Yao2011}. Because such representation enlarges the physical Hilbert space, the constraint $D_i=-i\gamma_i^x\gamma_i^y\gamma_i^z d_i^x d_i^y d_i^z=1$ needs to be enforced. In this representation, Eq. (\ref{OriginalH}) can be written in terms of the Majorana fermion operators
\begin{equation}
    H=\sum_{\langle i j\rangle,\alpha} iJ_{ij}u_{ij}\gamma_i^\alpha\gamma_j^\alpha
    +\frac{i\chi}{4}\sum_{\langle ij\rangle\langle jk\rangle,\alpha}\hat{u}_{ij}\hat{u}_{jk}\gamma_i^\alpha\gamma_k^\alpha,    
 \label{MajoranaH}
\end{equation}
where $u_{ij}=-id_i^\lambda d_j^\lambda$ and $J_{ij}=J_\lambda/4$ on the type-$\lambda$ link. Since $[H,u_{ij}]=0$ and $[u_{ij},u_{i^\prime j^\prime}]=0$, the set of bond variables $\{u_{ij}\}$ are good quantum numbers that have eigenvalues $\pm 1$ and the Hamiltonian Eq. (\ref{MajoranaH}) can be solved for each different $\{u_{ij}\}$. In fact, $u_{ij}$ acts as a $Z_2$ gauge field and $D_i$ serves as a generator of the $Z_2$ gauge symmetry. Gauge-invariant $Z_2$ flux operators can be defined on each plaquette $W_p=\prod_{\langle jk\rangle\in p} u_{jk}$ ($j\in A$ sublattice, $k\in B$ sublattice), which are good quantum numbers that label the physical eigenstates. 

Equation (\ref{MajoranaH}) can be regarded as three copies of the original Kitaev model. The global SO(3) symmetry among the three species of Majorana fermions originates from the original spin rotational symmetry. The phase diagram of Eq. (\ref{MajoranaH}) can hence be inferred from that of Kitaev model \cite{Kitaev2006}: when $\chi=0$ and $|J_x|$, $|J_y|$, $|J_z|$ satisfy the triangle inequalities, it describes a gapless QSOL; when $\chi=0$ and $|J_x|$, $|J_y|$, $|J_z|$ do not meet the triangle inequality conditions, it describes a non-chiral gapped QSOL; at the isotropic point $J_x=J_y=J_z$, if $\chi\neq 0$, it is a chiral gapped QSOL.

It is convenient to define complex fermion operators $f_{i,z}=(\gamma_i^x-i\gamma_i^y)/2$ and Eq. (\ref{MajoranaH}) can be rewritten as sum of two parts
\begin{equation}
    H=H_\text{K}+H_\text{H}=\sum_{ij}\frac{h_{ij}}{4}\gamma_{i}^z \gamma_{j}^z+\sum_{ij}h_{ij}f_{i,z}^\dagger f_{j,z} ,
 \label{magnonMajoranaH}
\end{equation}
where $H_\text{K}$ is the Kitaev model, while at the zero flux sector and isotropic point $H_\text{H}$ is equivalent to the Haldane model \cite{Haldane1988}. The matrix elements $h_{ij}$ when $u_{ij}=1$ are indicated in Fig. \ref{fig:lattice}. Because $S_i^z=f_{i,z}^\dagger f_{i,z}-\frac{1}{2}$, the fermions created by $f_{i,z}^\dagger$ carry $S_z=1$ and are hence dubbed as FMs.
 \begin{figure*}[t]
     \centering
     \includegraphics[height=1.6 in]{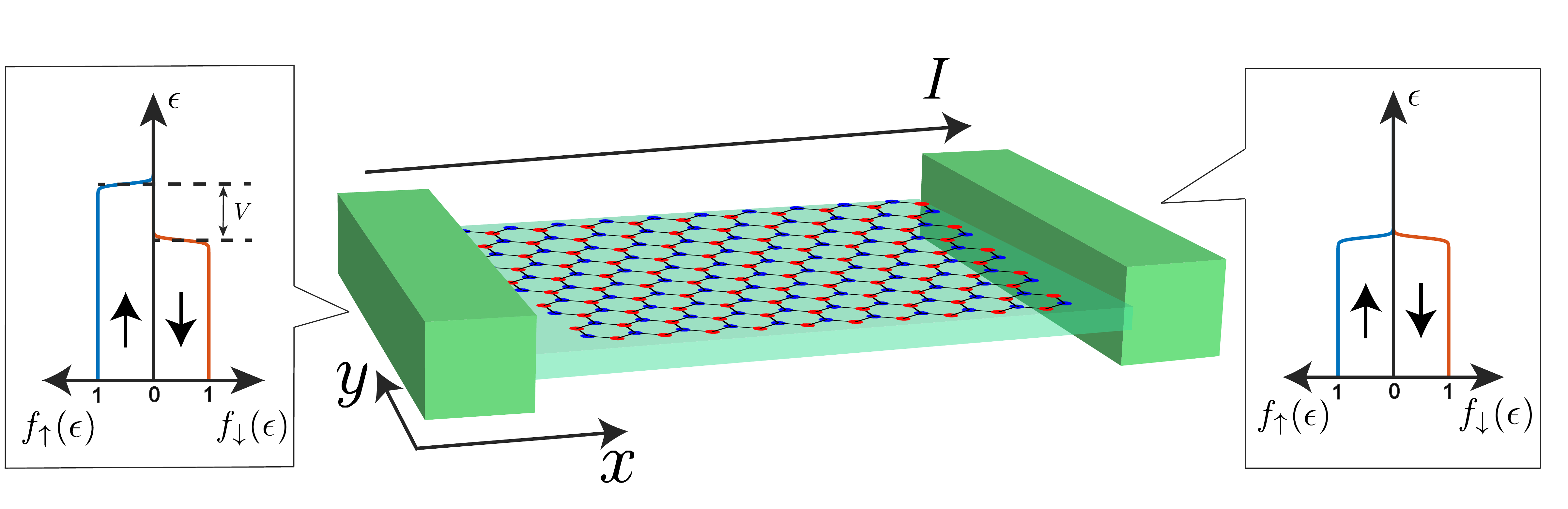}
     \caption{Schematic of the experiment setup. The spin-up electrons and spin-down electrons in the left spin bath have chemical potential difference $V_s$, which drives spin current $I_s$ through the structure.}
     \label{fig:setup}
 \end{figure*}

We consider an experimental setup as shown in Fig. \ref{fig:setup}, where the system is sandwiched between two spin baths. The spin baths are paramagnetic metals described by the Hamiltonian  $H^\alpha_{\text{bath}}=\sum_{\alpha,n}\epsilon_{\alpha n} c^{\dagger}_{\alpha n\sigma} c_{\alpha n\sigma}$, where $\sigma$ labels spin, $n$ labels eigenstate and $\alpha=L,R$ labels the bath. The system couples with the spin bath $\alpha$ through the Heisenberg exchange interaction at the boundary $A_\alpha$
\begin{equation}
H^\alpha_\text{int}=\sum_{\langle i,i_\alpha\rangle\in A_\alpha} \lambda^\alpha \vec{S_i}\cdot\vec{S}_{i_\alpha }
 \label{Hint}
\end{equation}
where $\vec{S_i}$ represents the spin $i$ in the system while $\vec{S}_{i_\alpha}$ denotes the spin $i_\alpha$ in the spin bath $\alpha$ that interacts with spin $i$. It is proposed that a spin bias $V_s=\mu_{\uparrow}-\mu_{\downarrow}$ can be induced in one bath, for example by the spin Hall effect, where $\mu_{{\uparrow}(\downarrow)}$ is the chemical potential of spin-up (down) electrons, while the spin current $I_s$ through the structure can be detected, for instance by inverse spin Hall effect, in the other bath \cite{Chatterjee2015,cornelissen2015,wesenberg2017,lebrun2018}. In this work, we will simply assume $V_L=V_s$, $V_R=0$ and spin current flows from left to right. We only consider the gapless phase and chiral gapped phase, as the non-chiral gapped phase generally does not have gapless excitations.

\begin{figure}[b]
  \subfigure[]{
    \label{fig:FOdiagram} %% label for second subfigure
    \includegraphics[width=0.45 \textwidth]{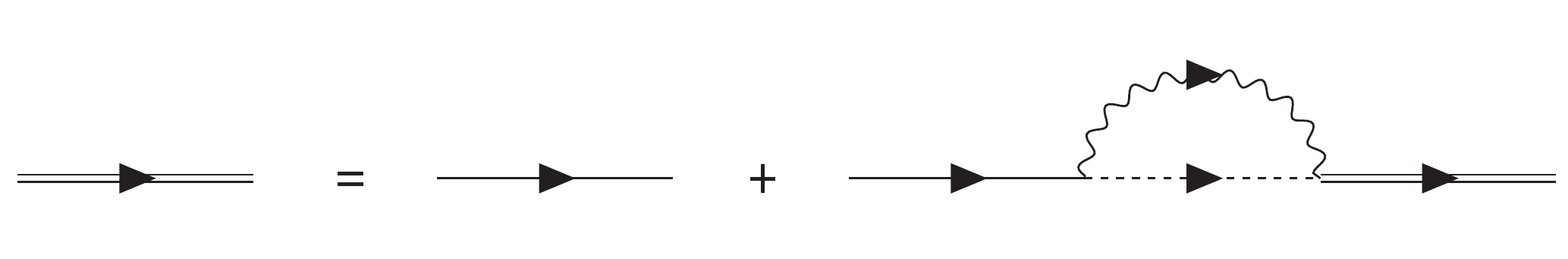}}
          \subfigure[]{
    \label{fig:SCdiagram} %% label for first subfigure
    \includegraphics[width=0.45 \textwidth]{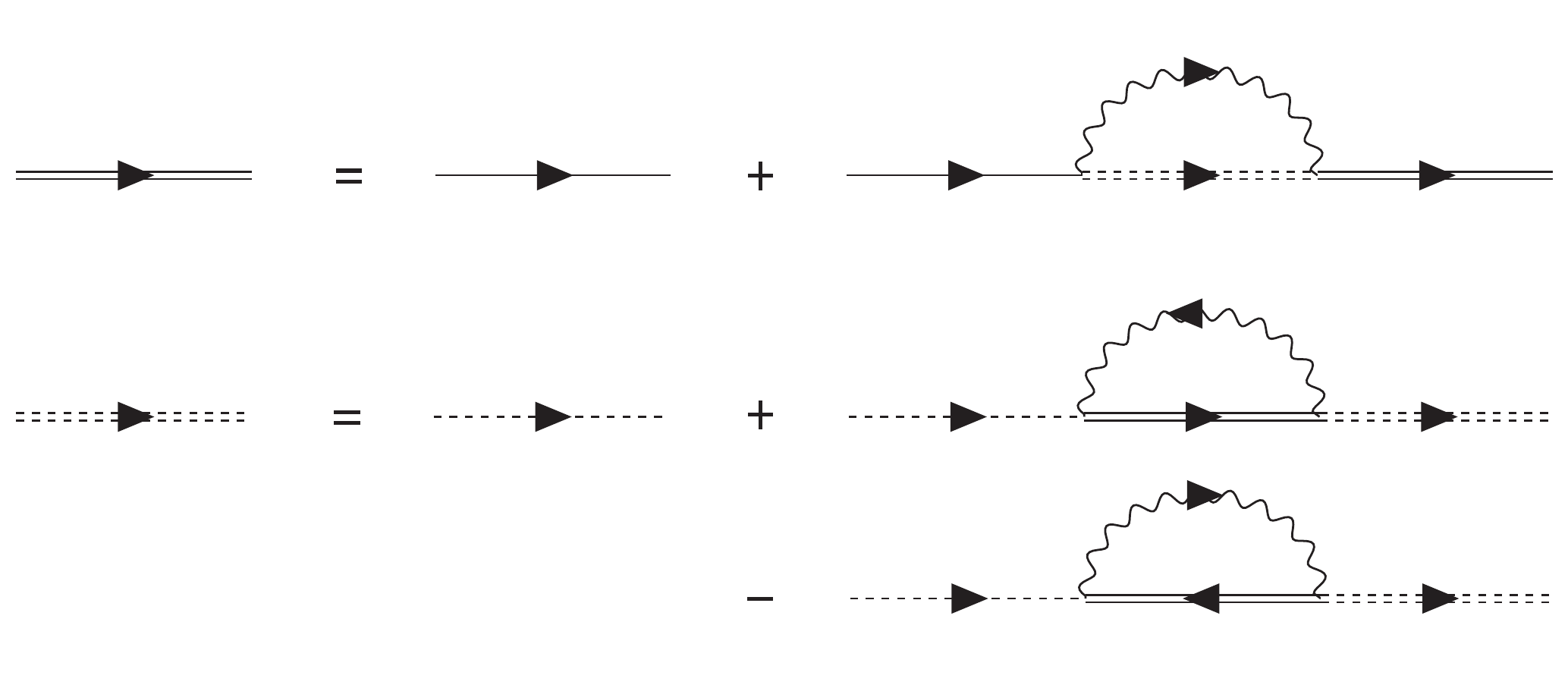}}
    \caption{The Dyson equations for FMs (solid lines) and Majorana fermions (dashed lines). The double lines indicate the full propagators while the single lines are for bare propagators. The wavy lines refer to $g^M$ defined in the text. The diagrams in (a) represent the calculation of $\Sigma^{f(1)}$ in Eq. (\ref{Sigf1}), while the diagrams of (b) are for the calculation of $\Sigma^{f(sc)}$ and $\Sigma^{\gamma(sc)}$ in Eq. (\ref{Sigf}) and Eq. (\ref{Signu}). }
\label{fig:diagrams}
\end{figure}

\textit{Formalism.}---We use NEGFs to investigate the spin transport in the forementioned model \cite{rammer2007,mahan2013}. The idea is similar to the calculation of electric current in a mesoscopic electronic system \cite{Caroli1971,Meir1992,Datta1997,Zhuang2020}. We aim to calculate gauge-invariant observables, which is given by
\begin{equation}
    \langle O(t)\rangle=\text{Tr} \left (\rho_{\text{init}}[U(t,-\infty)]^\dagger O U(t,-\infty) \right) \label{ExpOfO}
\end{equation}
where $U(t,-\infty)=e^{-i\int_{-\infty}^t   H_\text{int}(\tau)d\tau}$, $\rho _{\text{init}}= \rho_L\otimes\rho_S\otimes\rho_R$, $\rho_{L(R)}$ is the density matrix of the left(right) bath with  $\mu_{\uparrow}-\mu_{\downarrow}=V_{L(R)}$, $\rho_S=e^{-\beta H}/\text{Tr}(e^{-\beta H})$ is the density matrix of the system, all at temperature $T=1/\beta$. Note we use the interaction representation here. Equation (\ref{ExpOfO}) may be evaluated by representing all Pauli matrices with Majorana fermions. In this work, we assume that the flux gap $\Delta_{\text{flux}}$ is much larger than the temperature $T$ and the spin bias $V_s$ so that we only need to focus on the fluxless gauge sector, which has the lowest energy according to Lieb's theorem \cite{Lieb1994}. Because both the hybridization (\ref{Hint}) and the observable we are interested in, i.e. spin current, do not mix different gauge sectors, we will simply choose $u_{jk}=1$  and work within this gauge choice. With Wick's theorem one can therefore decompose Eq. (\ref{ExpOfO}) to products of Green's functions and evaluate it with Keldysh techniques \cite{Mao2003}. When $\chi = 0$ the spin current operator $I_{ij}$ is given by
\begin{equation}
\begin{split}
    I_{ij}&=\frac{iJ_\lambda}{2}[\tau_i^\lambda\tau_j^\lambda](\sigma_i^-\sigma_j^+-\sigma_i^+\sigma_j^-)\\
    &=2J_{ij}u_{ij}(f_{j,z}^\dagger f_{i,z}+f_{i,z}^\dagger f_{j,z}),\\
\end{split}
\end{equation}
where $i$ and $j$ are nearest-neighbor sites connected by $\lambda$-type bond and the spin current flows from $j$ to $i$. For convenience we set $\hbar=1$ throughout this paper. Thus one needs to calculate the full FM propagator $G^f_{ij}(t,t^\prime)=-i\langle \mathcal{T}_c e^{-i\int_c d\tau H_{\text{int}}(\tau)}f_{i,z}(t)f_{j,z}^\dagger(t^\prime)\rangle$, which can be approximately obtained by re-summing relevant diagrams and calculating the Dyson equation (see Supplemental Material \cite{SM}, Sec. I for more details)
\begin{equation}
    G_{ij}^{f}=g_{ij}^f+\sum_\alpha G_{ik}^f\Sigma^f_{\alpha,kl}g_{lj}^f \label{GfDyson}
\end{equation}
where $g^f_{ij}$ is the bare propagator for FMs. The convolution on the Keldysh contour and the sum over repeated indices have been implicitly indicated. Note that the time arguments of the Green's functions are incorporated in the subscripts when not written explicitly. The simplest self-energy $\Sigma^{f(1)}_{\alpha,ij}$ due to the bath $\alpha$ ($i,j\in A_\alpha$) is given by Fig. \ref{fig:FOdiagram}
 \begin{equation}
 \Sigma^{f(1)}_{\alpha,ij}=\frac{i(\lambda^\alpha)^2}{4}g^M_{\alpha, ij}g^\gamma_{ij} \label{Sigf1}    
 \end{equation}
while the self-consistent self-energy $\Sigma^{f(sc)}_{\alpha,ij}$  is (see Fig. \ref{fig:SCdiagram})
\begin{equation}
    \Sigma^{f(sc)}_{\alpha,ij}=\frac{i(\lambda^\alpha)^2}{4}g^M_{\alpha, ij}G^\gamma_{ij}, \label{Sigf}
\end{equation}
where $g^M_{\alpha,ij}(t,t^\prime)=-i\sum_{i_\alpha,j_\alpha}\langle \mathcal{T}_c S_{i_\alpha }^-(t)S_{j_\alpha}^+(t^\prime)\rangle$. In Eq. (\ref{Sigf}) the full Majorana propagator $G^\gamma_{ij}(t,t^\prime)=-i\langle \mathcal{T}_c e^{-i\int_c d\tau H_{\text{int}}(\tau)}\gamma_i^z(t)\gamma_j^z(t^\prime)\rangle$ satisfies Dyson equation
\begin{equation}
    G_{ij}^{\gamma}=g_{ij}^\gamma+\sum_\alpha G_{ik}^\gamma\Sigma^{\gamma(sc)}_{\alpha,kl} g_{lj}^\gamma \label{GnuDyson}
\end{equation}
where $g_{ij}^\gamma(t,t^\prime)$ is the bare Majorana propagator and the self-energy $\Sigma^{\gamma(sc)}_{\alpha,ij}$ ($i,j\in A_\alpha$) is given by 
\begin{equation}
    \Sigma^{\gamma(sc)}_{\alpha,ij}=\frac{i(\lambda^\alpha)^2}{4}\left[g^M_{\alpha,ji}G^f_{ij}-g^M_{\alpha,ij}G^f_{ji}\right].
\label{Signu}
\end{equation}
Equation (\ref{GfDyson}) and (\ref{Sigf1}) give a first-order solution while the closed set of Eqs. (\ref{GfDyson})(\ref{Sigf})(\ref{GnuDyson})(\ref{Signu}) can be solved iteratively to obtain a self-consistent solution. We point out that one class of diagrams that we neglect corresponds to the $S^z_i S^z_{i_\alpha}$ term as it does not lead to dissipation and only contributes to the real part of the self-energy, slightly renormalizing the Hamiltonian. The term does not alter the transport qualitatively. Although throwing away such terms may break the original SU(2) spin rotational symmetry, the U(1) charge $S_z$ is still conserved and hence the spin current along the $z$ direction is still well defined. The other neglected diagrams include those with dressed vertices, dressed $g^M$, and diagrams that cannot be represented in terms of $g^M$.

From the perspective of NEGFs, spin transport in Yao-Lee model and the electron transport in graphene are similar, not only because both the Hamiltonian and current operators are similar, but also because the non-equilibrium dynamics is determined by the self-energies at the system-bath interface. For FMs the effects of Majorana fermions $\gamma_z$ and spin bath $\alpha$ are equivalent to that of an effective fermionic bath with spectral function
\begin{equation}
    \bm{\Gamma}^\alpha(\omega)=-\frac{\bm{\Sigma}^{f,R}_\alpha(\omega)-\bm{\Sigma}^{f,A}_\alpha(\omega)}{2\pi i } \label{effDOS}
\end{equation} and distribution function 
\begin{equation}
    f^\alpha_{\text{eff}}(\omega)=\frac{1}{2}\left(1-\frac{\Sigma^{f,K}_{\alpha,ij}(\omega)}{\Sigma^{f,R}_{\alpha,ij}(\omega)-\Sigma^{f,A}_{\alpha,ij}(\omega)}\right). \label{disexpression}
\end{equation}
if a single well-defined distribution function exists. The superscripts $R,A,K$ denote the retarded, advanced and Keldysh components respectively, and bold symbols represent matrices. We will refer such effective fermionic bath as `FM bath' in this work. It can be shown that the distribution function calculated by $\Sigma^{f(1)}_\alpha$ is (see Supplemental Material \cite{SM}, Sec. II)
\begin{equation}
    f^{\alpha(1)}_{\text{eff}}(\omega)=\frac{1}{e^{\beta(\omega-V_\alpha)}+1}, \label{disBath}
\end{equation}
which indicates that the FM bath $\alpha$ has exactly chemical potential $V_\alpha$.
To calculate the spectral function explicitly, we make local self-energy approximation (LSEA) which assumes that the self-energy is local in space. We note that this approximation is not essential and does not alter the calculation qualitatively as long as the self-energies in real space all have same $\omega$-dependence at low energies. Within this approximation the spectral function at small $\omega$ is given by 
\begin{multline}
    \Gamma^{\alpha(1)}_{ii}(\omega+V_\alpha)
   \approx\frac{(\lambda^\alpha J^\alpha)^2}{4} \int d\omega^\prime  \omega^\prime D^\alpha_{ii}(\omega-\omega^\prime)\\\times\left[\tanh \frac{\beta(\omega-\omega^\prime)}{2}+\coth \frac{\beta \omega^\prime}{2}\right] , \label{selfEsimplify}
\end{multline}
where $J^\alpha$ is the local density of state (LDOS) per spin (of spin bath $\alpha$ at the interface $A_\alpha$) and $D^\alpha_{ii}=-\text{Im}g^{\gamma,R}_{i i}/2\pi$ ($i \in A_\alpha$).

The self-consistent self-energy  $\Sigma^{f(sc)}_{\alpha}$ generally may give a more complicated correction to both spectral functions and distribution functions. However, within LSEA, we numerically find that it only gives a minor correction to the spectral function $\bm{\Gamma}$ and changes the FM bath temperature $T=1/\beta$ in $f^{\alpha(1)}_\text{eff}$ to an effective temperature $\tilde{T}^\alpha=1/\beta^\alpha$, if the coupling $\lambda^\alpha$ is not too strong (see Supplemental Material \cite{SM}, Sec. III). Therefore we will assume the effects of $\Sigma^{f(sc)}_{\alpha}$ is negligible and use $\Sigma^{f(1)}_{\alpha}$ to investigate the transport in the gapless phase. Since the spin transport problem has been mapped to a fermion transport problem, we directly apply the Meir-Wingreen formula for non-interacting fermions \cite{Meir1992}
\begin{equation}
    I_s=2\pi\int d\omega\left[ f^L_{\text{eff}}(\omega)-f^R_{\text{eff}}(\omega)\right] \text{Tr}\left(\bm{G}^{f,A}\bm{\Gamma}^R_{\text{eff}}\bm{G}^{f,R}\bm{\Gamma}^L_{\text{eff}}\right) \label{MWformula}
\end{equation}
to obtain the total spin current passing through the structure in the gapless phase. Note that in this work when calculating $G^{f,R(A)}$ numerically we ignore the real part of self-energy and assume it does no affect our final results significantly.
\begin{figure}

    \subfigure[]{
    \label{subfig:armchair}
    \includegraphics[height=1.3 in]{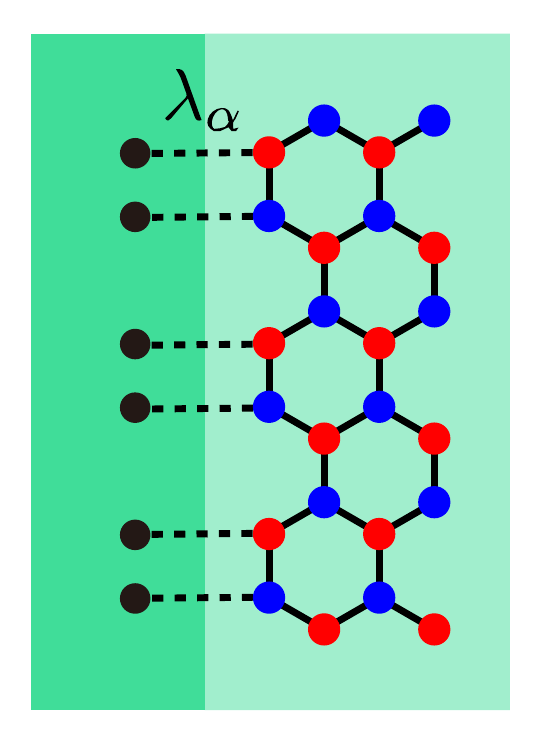}
    }    
    ~
    \subfigure[]{
    \label{subfig:zigzag}
    \includegraphics[height=1.3 in]{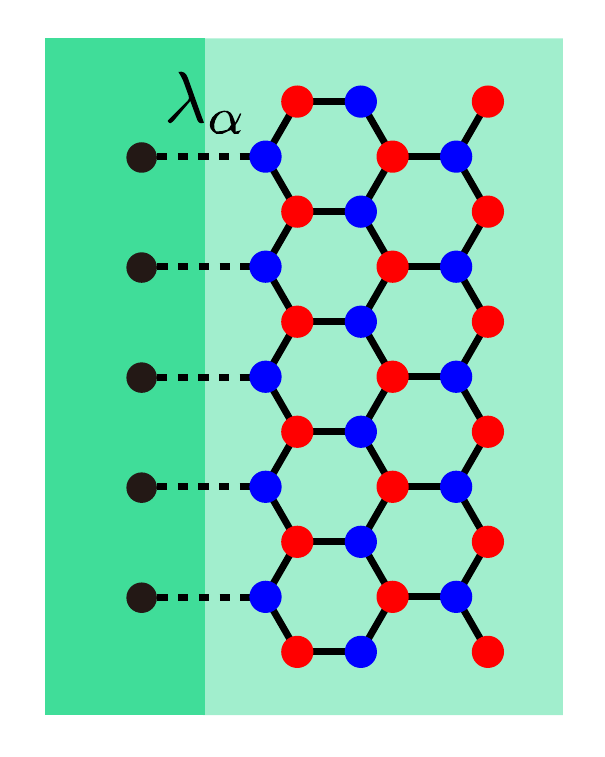}
    }
    \caption{The (a) armchair-type and (b) zigzag-type contacts. Dashed lines represent the Heisenberg interaction and black dots denote the sites in the spin baths.}
    \label{fig:contact}
\end{figure}

\textit{Gapless phase.}---We impose periodic boundary condition along the $y$-direction and focus on the isotropic point $J_x=J_y=J_z$ for simplicity. If the width $L_y$ is finite, series of FM subbands labelled by discrete $k_y$ are developed. We assume that $V_\alpha\gg v_F/L_y$, where $v_F$ is the Fermi velocity at Dirac point. This condition indicates that there are many transverse modes participating the spin transport. The spectral function of FM bath is rapidly varying: for the armchair-type contact (AC) depicted in Fig. \ref{subfig:armchair}, $D(\omega)\sim |\omega|$, therefore according to Eq. (\ref{selfEsimplify}) $\Gamma^{\alpha(1)}(\omega+V_\alpha)\sim |\omega|^3$ when $\omega \gg T$; for the zigzag-type contact (ZC) shown in Fig. \ref{subfig:zigzag}, due to the existence of localized Majorana zero modes at boundaries \cite{Kohmoto2007,Thakurathi2014,DeCarvalho2018,Mizoguchi2019}, $D(\omega)\sim \delta(\omega)$ so $\Gamma^{\alpha(1)}_{\text{eff}}(\omega+V_\alpha)\sim |\omega|$ if $\omega \gg T$. We show below that when $V_s\gg T$, $I_s\sim V_s^5$ for AC and $I_s\sim V_s^3$ for ZC. For both types of contact, $I_s\sim V_s$ when $V_s\ll T$. We have verified these power laws by evaluating Eq. (\ref{MWformula}) numerically (see Supplemental Material \cite{SM}, Sec. V). The spin current does not significantly depend on the system's length $L_x$ if the system is sufficiently long.

\textit{Chiral gapped phase.}---We first consider an infinitely long edge of Yao-Lee model in the chiral gapped phase, connected with a single spin bath $\alpha$. In this phase the bulk is gapped and on the edge there are three chiral Majorana modes $\gamma^\alpha$, or equivalently one chiral FM mode $f_z$ plus one chiral Majorana mode $\gamma^z$.As the edge is only connected with one spin bath, calculation of $\Sigma^{f(1)}$ and $\Sigma^{f(sc)}$ both give $f^{\alpha(1)}_{\text{eff}}(\omega)=f^{\alpha(sc)}_{\text{eff}}(\omega)=\frac{1}{e^{\beta(\omega-V_\alpha)}+1}$ (see Supplemental Material \cite{SM}, Sec. III). This indicates that even for a finite-size system with open boundary condition, the chiral FMs near the contact $\alpha$ also have a well-defined temperature $T$ and a chemical potential $V_\alpha$, as long as the contact length $L_y$ satisfy condition
\begin{equation}
    L_y\gg \frac{v_C}{\text{min}\{|\text{Im} \Sigma_{ii}^{f,R}(\omega)|\}}\sim \frac{v_C^2}{a(\lambda^\alpha J^\alpha)^2T^2}
\end{equation}
where $v_C$ is the Fermi velocity of the chiral modes, $a$ is the bond length defined in Fig. \ref{fig:lattice}, and we have used Eq. (\ref{selfEsimplify}). Due to the chirality of the FM on the edge, the FMs carry the same distribution function after they leave the spin bath $\alpha$ until they reach the other spin bath, if there is no inelastic scattering or backscattering across the bulk. The spin current in the whole system is hence quantized to
\begin{equation}
    I_s=\frac{1}{2\pi}V_s,
\end{equation}
similar to that in the integer quantum Hall effect or quantum anomalous Hall effect.

\textit{Discussion.}---There are similarities and differences between our results and those  reported in Ref. \cite{DeCarvalho2018}. We first remark that the spin currents found using equilibrium spin correlation functions \cite{Chen2013,Chatterjee2015,DeCarvalho2018}
are equivalent to a calculation of the tunneling current between the left FM bath and the FM honeycomb model at equilibrium (see Supplemental Material \cite{SM}, Section IV). This indicates that at zero temperature $I_s\sim \int_0^{V_\alpha} d\omega N(\omega)\Gamma^{L(1)}(\omega)$, where $N(\omega)$ is LDOS of FM sector at the left interface. Therefore $I_s\sim V_s^5$ ($V_s$) for the gapless phase with ACs (ZCs), and $I_s\sim V_s^3$ for the chiral gapped phase are obtained. The expressions correctly capture the power law $I_s-V_s$ relations for the gapless phase with ACs, and as discussed below, may also qualitatively explain the power law for the gapless phase with ZCs if certain subtleties are taken into consideration. Nevertheless, the above formula cannot be applied to the chiral gapped phase when the contact is sufficiently wide, as the FMs at the contact are highly out of equilibrium. We expect, and have numerically verified, that the scaling relation $I_s\sim V_s^3$ can be restored when the contact is narrow enough.

For the gapless phase with ZCs, the predicted linear $I_s-V_s$ relation originates from the dominant zero-frequency peak of LDOS at the zigzag edge \cite{CastroNeto2009,DasSarma2011}. However, in a sufficiently clean and long system, these FM zero modes do not play a role in transport as hopping between modes on opposing edges is exponentially suppressed with increasing separation. In other words, for positive $V_\alpha$ the FM zero modes at the left edge would be fully occupied, and FMs cannot tunnel from the left FM bath to these modes. A qualitatively correct power law may still be obtained using the above formula, if one neglects the contribution from these FM zero modes and uses $N(\omega) \sim |\omega|$ instead. Our results indicate that attention needs to be paid to non-equilibrium physics to fully understand spin transport.

\textit{Conclusion.}---In this Letter we use NEGFs to describe the non-equilibrium spin transport in the Yao-Lee QSOL model. Our results regarding the gapless phase with ZCs ($I_s \sim V_s^3$) and chiral gapped phase ($I_s=V_s/2\pi$) are different from those obtained in earlier work, showing the importance of the non-equilibrium physics. The quantized spin current conductance can test for the existence of chiral FMs on the boundary of a topological QSOL. It is an open and interesting question as to how our results would be modified by the inclusion of neglected diagrams as well as interactions that move the QSOL away from exactly solvable limit. Our work paves the way to understand spin transport in QSOLs over a broader parameter range and in the presence of disorder and thermally excited fluxes.

We thank C.-Z. Chen and S. Chatterjee for useful email discussions. We are grateful to K. Plumb for enlightening discussions and references. We also thank D. E. Feldman, J. Merino and V. F. Mitrovi{\'c} for reading the manuscript and providing helpful feedback. This work was supported in part by U.S. National Science Foundation grants OIA-1921199 and OMA-1936221.

\bibliography{Reference}

\begin{thebibliography}{62}%
\makeatletter
\providecommand \@ifxundefined [1]{%
 \@ifx{#1\undefined}
}%
\providecommand \@ifnum [1]{%
 \ifnum #1\expandafter \@firstoftwo
 \else \expandafter \@secondoftwo
 \fi
}%
\providecommand \@ifx [1]{%
 \ifx #1\expandafter \@firstoftwo
 \else \expandafter \@secondoftwo
 \fi
}%
\providecommand \natexlab [1]{#1}%
\providecommand \enquote  [1]{``#1''}%
\providecommand \bibnamefont  [1]{#1}%
\providecommand \bibfnamefont [1]{#1}%
\providecommand \citenamefont [1]{#1}%
\providecommand \href@noop [0]{\@secondoftwo}%
\providecommand \href [0]{\begingroup \@sanitize@url \@href}%
\providecommand \@href[1]{\@@startlink{#1}\@@href}%
\providecommand \@@href[1]{\endgroup#1\@@endlink}%
\providecommand \@sanitize@url [0]{\catcode `\\12\catcode `\$12\catcode
  `\&12\catcode `\#12\catcode `\^12\catcode `\_12\catcode `\%12\relax}%
\providecommand \@@startlink[1]{}%
\providecommand \@@endlink[0]{}%
\providecommand \url  [0]{\begingroup\@sanitize@url \@url }%
\providecommand \@url [1]{\endgroup\@href {#1}{\urlprefix }}%
\providecommand \urlprefix  [0]{URL }%
\providecommand \Eprint [0]{\href }%
\providecommand \doibase [0]{https://doi.org/}%
\providecommand \selectlanguage [0]{\@gobble}%
\providecommand \bibinfo  [0]{\@secondoftwo}%
\providecommand \bibfield  [0]{\@secondoftwo}%
\providecommand \translation [1]{[#1]}%
\providecommand \BibitemOpen [0]{}%
\providecommand \bibitemStop [0]{}%
\providecommand \bibitemNoStop [0]{.\EOS\space}%
\providecommand \EOS [0]{\spacefactor3000\relax}%
\providecommand \BibitemShut  [1]{\csname bibitem#1\endcsname}%
\let\auto@bib@innerbib\@empty
%</preamble>
\bibitem [{\citenamefont {Savary}\ and\ \citenamefont
  {Balents}(2016)}]{Savary2016}%
  \BibitemOpen
  \bibfield  {author} {\bibinfo {author} {\bibfnamefont {L.}~\bibnamefont
  {Savary}}\ and\ \bibinfo {author} {\bibfnamefont {L.}~\bibnamefont
  {Balents}},\ }\bibfield  {title} {\bibinfo {title} {Quantum spin liquids: a
  review},\ }\href {https://doi.org/10.1088/0034-4885/80/1/016502} {\bibfield
  {journal} {\bibinfo  {journal} {Reports on Progress in Physics}\ }\textbf
  {\bibinfo {volume} {80}},\ \bibinfo {pages} {016502} (\bibinfo {year}
  {2016})}\BibitemShut {NoStop}%
\bibitem [{\citenamefont {Zhou}\ \emph {et~al.}(2017)\citenamefont {Zhou},
  \citenamefont {Kanoda},\ and\ \citenamefont {Ng}}]{Zhou2017}%
  \BibitemOpen
  \bibfield  {author} {\bibinfo {author} {\bibfnamefont {Y.}~\bibnamefont
  {Zhou}}, \bibinfo {author} {\bibfnamefont {K.}~\bibnamefont {Kanoda}},\ and\
  \bibinfo {author} {\bibfnamefont {T.-K.}\ \bibnamefont {Ng}},\ }\bibfield
  {title} {\bibinfo {title} {Quantum spin liquid states},\ }\href
  {https://doi.org/10.1103/RevModPhys.89.025003} {\bibfield  {journal}
  {\bibinfo  {journal} {Rev. Mod. Phys.}\ }\textbf {\bibinfo {volume} {89}},\
  \bibinfo {pages} {025003} (\bibinfo {year} {2017})}\BibitemShut {NoStop}%
\bibitem [{\citenamefont {Broholm}\ \emph {et~al.}(2020)\citenamefont
  {Broholm}, \citenamefont {Cava}, \citenamefont {Kivelson}, \citenamefont
  {Nocera}, \citenamefont {Norman},\ and\ \citenamefont
  {Senthil}}]{broholm2020}%
  \BibitemOpen
  \bibfield  {author} {\bibinfo {author} {\bibfnamefont {C.}~\bibnamefont
  {Broholm}}, \bibinfo {author} {\bibfnamefont {R.}~\bibnamefont {Cava}},
  \bibinfo {author} {\bibfnamefont {S.}~\bibnamefont {Kivelson}}, \bibinfo
  {author} {\bibfnamefont {D.}~\bibnamefont {Nocera}}, \bibinfo {author}
  {\bibfnamefont {M.}~\bibnamefont {Norman}},\ and\ \bibinfo {author}
  {\bibfnamefont {T.}~\bibnamefont {Senthil}},\ }\bibfield  {title} {\bibinfo
  {title} {Quantum spin liquids},\ }\href
  {https://science.sciencemag.org/content/367/6475/eaay0668} {\bibfield
  {journal} {\bibinfo  {journal} {Science}\ }\textbf {\bibinfo {volume} {367}}
  (\bibinfo {year} {2020})}\BibitemShut {NoStop}%
\bibitem [{\citenamefont {Kitaev}(2006)}]{Kitaev2006}%
  \BibitemOpen
  \bibfield  {author} {\bibinfo {author} {\bibfnamefont {A.}~\bibnamefont
  {Kitaev}},\ }\bibfield  {title} {\bibinfo {title} {Anyons in an exactly
  solved model and beyond},\ }\href
  {https://doi.org/https://doi.org/10.1016/j.aop.2005.10.005} {\bibfield
  {journal} {\bibinfo  {journal} {Annals of Physics}\ }\textbf {\bibinfo
  {volume} {321}},\ \bibinfo {pages} {2 } (\bibinfo {year} {2006})}\BibitemShut
  {NoStop}%
\bibitem [{\citenamefont {Jackeli}\ and\ \citenamefont
  {Khaliullin}(2009)}]{Jackeli2009}%
  \BibitemOpen
  \bibfield  {author} {\bibinfo {author} {\bibfnamefont {G.}~\bibnamefont
  {Jackeli}}\ and\ \bibinfo {author} {\bibfnamefont {G.}~\bibnamefont
  {Khaliullin}},\ }\bibfield  {title} {\bibinfo {title} {{Mott} insulators in
  the strong spin-orbit coupling limit: {From} {Heisenberg} to a quantum
  compass and {Kitaev} models},\ }\href
  {https://doi.org/10.1103/PhysRevLett.102.017205} {\bibfield  {journal}
  {\bibinfo  {journal} {Phys. Rev. Lett.}\ }\textbf {\bibinfo {volume} {102}},\
  \bibinfo {pages} {017205} (\bibinfo {year} {2009})}\BibitemShut {NoStop}%
\bibitem [{\citenamefont {Kugel}\ and\ \citenamefont
  {Khomskii}(1973)}]{kugel1973}%
  \BibitemOpen
  \bibfield  {author} {\bibinfo {author} {\bibfnamefont {K.}~\bibnamefont
  {Kugel}}\ and\ \bibinfo {author} {\bibfnamefont {D.}~\bibnamefont
  {Khomskii}},\ }\bibfield  {title} {\bibinfo {title} {Crystal-structure and
  magnetic properties of substances with orbital degeneracy},\ }\href
  {http://jetp.ac.ru/cgi-bin/dn/e_037_04_0725.pdf} {\bibfield  {journal}
  {\bibinfo  {journal} {Zh. Eksp. Teor. Fiz}\ }\textbf {\bibinfo {volume}
  {64}},\ \bibinfo {pages} {1429} (\bibinfo {year} {1973})}\BibitemShut
  {NoStop}%
\bibitem [{\citenamefont {Kugel}\ and\ \citenamefont
  {Khomski{\u\i}}(1982)}]{kugel1982}%
  \BibitemOpen
  \bibfield  {author} {\bibinfo {author} {\bibfnamefont {K.~I.}\ \bibnamefont
  {Kugel}}\ and\ \bibinfo {author} {\bibfnamefont {D.}~\bibnamefont
  {Khomski{\u\i}}},\ }\bibfield  {title} {\bibinfo {title} {{The Jahn-Teller
  effect and magnetism: transition metal compounds}},\ }\href
  {https://doi.org/10.1070/PU1982v025n04ABEH004537} {\bibfield  {journal}
  {\bibinfo  {journal} {Soviet Physics Uspekhi}\ }\textbf {\bibinfo {volume}
  {25}},\ \bibinfo {pages} {231} (\bibinfo {year} {1982})}\BibitemShut
  {NoStop}%
\bibitem [{\citenamefont {Yao}\ and\ \citenamefont {Lee}(2011)}]{Yao2011}%
  \BibitemOpen
  \bibfield  {author} {\bibinfo {author} {\bibfnamefont {H.}~\bibnamefont
  {Yao}}\ and\ \bibinfo {author} {\bibfnamefont {D.-H.}\ \bibnamefont {Lee}},\
  }\bibfield  {title} {\bibinfo {title} {Fermionic magnons, non-abelian
  spinons, and the spin quantum {Hall} effect from an exactly solvable
  spin-$1/2$ {Kitaev} model with {SU(2)} symmetry},\ }\href
  {https://doi.org/10.1103/PhysRevLett.107.087205} {\bibfield  {journal}
  {\bibinfo  {journal} {Phys. Rev. Lett.}\ }\textbf {\bibinfo {volume} {107}},\
  \bibinfo {pages} {087205} (\bibinfo {year} {2011})}\BibitemShut {NoStop}%
\bibitem [{\citenamefont {Feiner}\ \emph {et~al.}(1997)\citenamefont {Feiner},
  \citenamefont {Ole\ifmmode~\acute{s}\else \'{s}\fi{}},\ and\ \citenamefont
  {Zaanen}}]{feiner1997}%
  \BibitemOpen
  \bibfield  {author} {\bibinfo {author} {\bibfnamefont {L.~F.}\ \bibnamefont
  {Feiner}}, \bibinfo {author} {\bibfnamefont {A.~M.}\ \bibnamefont
  {Ole\ifmmode~\acute{s}\else \'{s}\fi{}}},\ and\ \bibinfo {author}
  {\bibfnamefont {J.}~\bibnamefont {Zaanen}},\ }\bibfield  {title} {\bibinfo
  {title} {Quantum melting of magnetic order due to orbital fluctuations},\
  }\href {https://doi.org/10.1103/PhysRevLett.78.2799} {\bibfield  {journal}
  {\bibinfo  {journal} {Phys. Rev. Lett.}\ }\textbf {\bibinfo {volume} {78}},\
  \bibinfo {pages} {2799} (\bibinfo {year} {1997})}\BibitemShut {NoStop}%
\bibitem [{\citenamefont {Ole\ifmmode~\acute{s}\else \'{s}\fi{}}\ \emph
  {et~al.}(2000)\citenamefont {Ole\ifmmode~\acute{s}\else \'{s}\fi{}},
  \citenamefont {Feiner},\ and\ \citenamefont {Zaanen}}]{oles2000}%
  \BibitemOpen
  \bibfield  {author} {\bibinfo {author} {\bibfnamefont {A.~M.}\ \bibnamefont
  {Ole\ifmmode~\acute{s}\else \'{s}\fi{}}}, \bibinfo {author} {\bibfnamefont
  {L.~F.}\ \bibnamefont {Feiner}},\ and\ \bibinfo {author} {\bibfnamefont
  {J.}~\bibnamefont {Zaanen}},\ }\bibfield  {title} {\bibinfo {title} {Quantum
  melting of magnetic long-range order near orbital degeneracy: Classical
  phases and {Gaussian} fluctuations},\ }\href
  {https://doi.org/10.1103/PhysRevB.61.6257} {\bibfield  {journal} {\bibinfo
  {journal} {Phys. Rev. B}\ }\textbf {\bibinfo {volume} {61}},\ \bibinfo
  {pages} {6257} (\bibinfo {year} {2000})}\BibitemShut {NoStop}%
\bibitem [{\citenamefont {Yao}\ \emph {et~al.}(2009)\citenamefont {Yao},
  \citenamefont {Zhang},\ and\ \citenamefont {Kivelson}}]{Yao2009}%
  \BibitemOpen
  \bibfield  {author} {\bibinfo {author} {\bibfnamefont {H.}~\bibnamefont
  {Yao}}, \bibinfo {author} {\bibfnamefont {S.-C.}\ \bibnamefont {Zhang}},\
  and\ \bibinfo {author} {\bibfnamefont {S.~A.}\ \bibnamefont {Kivelson}},\
  }\bibfield  {title} {\bibinfo {title} {Algebraic spin liquid in an exactly
  solvable spin model},\ }\href
  {https://doi.org/10.1103/PhysRevLett.102.217202} {\bibfield  {journal}
  {\bibinfo  {journal} {Phys. Rev. Lett.}\ }\textbf {\bibinfo {volume} {102}},\
  \bibinfo {pages} {217202} (\bibinfo {year} {2009})}\BibitemShut {NoStop}%
\bibitem [{\citenamefont {Corboz}\ \emph {et~al.}(2012)\citenamefont {Corboz},
  \citenamefont {Lajk\'o}, \citenamefont {L\"auchli}, \citenamefont {Penc},\
  and\ \citenamefont {Mila}}]{Corboz2012}%
  \BibitemOpen
  \bibfield  {author} {\bibinfo {author} {\bibfnamefont {P.}~\bibnamefont
  {Corboz}}, \bibinfo {author} {\bibfnamefont {M.}~\bibnamefont {Lajk\'o}},
  \bibinfo {author} {\bibfnamefont {A.~M.}\ \bibnamefont {L\"auchli}}, \bibinfo
  {author} {\bibfnamefont {K.}~\bibnamefont {Penc}},\ and\ \bibinfo {author}
  {\bibfnamefont {F.}~\bibnamefont {Mila}},\ }\bibfield  {title} {\bibinfo
  {title} {Spin-orbital quantum liquid on the honeycomb lattice},\ }\href
  {https://doi.org/10.1103/PhysRevX.2.041013} {\bibfield  {journal} {\bibinfo
  {journal} {Phys. Rev. X}\ }\textbf {\bibinfo {volume} {2}},\ \bibinfo {pages}
  {041013} (\bibinfo {year} {2012})}\BibitemShut {NoStop}%
\bibitem [{\citenamefont {Nakai}\ \emph {et~al.}(2012)\citenamefont {Nakai},
  \citenamefont {Ryu},\ and\ \citenamefont {Furusaki}}]{Nakai2012}%
  \BibitemOpen
  \bibfield  {author} {\bibinfo {author} {\bibfnamefont {R.}~\bibnamefont
  {Nakai}}, \bibinfo {author} {\bibfnamefont {S.}~\bibnamefont {Ryu}},\ and\
  \bibinfo {author} {\bibfnamefont {A.}~\bibnamefont {Furusaki}},\ }\bibfield
  {title} {\bibinfo {title} {{Time-reversal symmetric Kitaev model and
  topological superconductor in two dimensions}},\ }\href
  {https://doi.org/10.1103/PhysRevB.85.155119} {\bibfield  {journal} {\bibinfo
  {journal} {Phys. Rev. B}\ }\textbf {\bibinfo {volume} {85}},\ \bibinfo
  {pages} {155119} (\bibinfo {year} {2012})}\BibitemShut {NoStop}%
\bibitem [{\citenamefont {Nussinov}\ and\ \citenamefont {van~den
  Brink}(2015)}]{nussinov2015}%
  \BibitemOpen
  \bibfield  {author} {\bibinfo {author} {\bibfnamefont {Z.}~\bibnamefont
  {Nussinov}}\ and\ \bibinfo {author} {\bibfnamefont {J.}~\bibnamefont {van~den
  Brink}},\ }\bibfield  {title} {\bibinfo {title} {Compass models: Theory and
  physical motivations},\ }\href {https://doi.org/10.1103/RevModPhys.87.1}
  {\bibfield  {journal} {\bibinfo  {journal} {Rev. Mod. Phys.}\ }\textbf
  {\bibinfo {volume} {87}},\ \bibinfo {pages} {1} (\bibinfo {year}
  {2015})}\BibitemShut {NoStop}%
\bibitem [{\citenamefont {Natori}\ \emph {et~al.}(2016)\citenamefont {Natori},
  \citenamefont {Andrade}, \citenamefont {Miranda},\ and\ \citenamefont
  {Pereira}}]{Natori2016}%
  \BibitemOpen
  \bibfield  {author} {\bibinfo {author} {\bibfnamefont {W.~M.~H.}\
  \bibnamefont {Natori}}, \bibinfo {author} {\bibfnamefont {E.~C.}\
  \bibnamefont {Andrade}}, \bibinfo {author} {\bibfnamefont {E.}~\bibnamefont
  {Miranda}},\ and\ \bibinfo {author} {\bibfnamefont {R.~G.}\ \bibnamefont
  {Pereira}},\ }\bibfield  {title} {\bibinfo {title} {Chiral spin-orbital
  liquids with nodal lines},\ }\href
  {https://doi.org/10.1103/PhysRevLett.117.017204} {\bibfield  {journal}
  {\bibinfo  {journal} {Phys. Rev. Lett.}\ }\textbf {\bibinfo {volume} {117}},\
  \bibinfo {pages} {017204} (\bibinfo {year} {2016})}\BibitemShut {NoStop}%
\bibitem [{\citenamefont {Natori}\ \emph {et~al.}(2018)\citenamefont {Natori},
  \citenamefont {Andrade},\ and\ \citenamefont {Pereira}}]{Natori2018}%
  \BibitemOpen
  \bibfield  {author} {\bibinfo {author} {\bibfnamefont {W.~M.~H.}\
  \bibnamefont {Natori}}, \bibinfo {author} {\bibfnamefont {E.~C.}\
  \bibnamefont {Andrade}},\ and\ \bibinfo {author} {\bibfnamefont {R.~G.}\
  \bibnamefont {Pereira}},\ }\bibfield  {title} {\bibinfo {title}
  {{SU(4)}-symmetric spin-orbital liquids on the hyperhoneycomb lattice},\
  }\href {https://doi.org/10.1103/PhysRevB.98.195113} {\bibfield  {journal}
  {\bibinfo  {journal} {Phys. Rev. B}\ }\textbf {\bibinfo {volume} {98}},\
  \bibinfo {pages} {195113} (\bibinfo {year} {2018})}\BibitemShut {NoStop}%
\bibitem [{\citenamefont {Natori}\ \emph {et~al.}(2019)\citenamefont {Natori},
  \citenamefont {Nutakki}, \citenamefont {Pereira},\ and\ \citenamefont
  {Andrade}}]{Natori2019}%
  \BibitemOpen
  \bibfield  {author} {\bibinfo {author} {\bibfnamefont {W.~M.~H.}\
  \bibnamefont {Natori}}, \bibinfo {author} {\bibfnamefont {R.}~\bibnamefont
  {Nutakki}}, \bibinfo {author} {\bibfnamefont {R.~G.}\ \bibnamefont
  {Pereira}},\ and\ \bibinfo {author} {\bibfnamefont {E.~C.}\ \bibnamefont
  {Andrade}},\ }\bibfield  {title} {\bibinfo {title} {{SU(4) Heisenberg model
  on the honeycomb lattice with exchange-frustrated perturbations: Implications
  for twistronics and Mott insulators}},\ }\href
  {https://doi.org/10.1103/PhysRevB.100.205131} {\bibfield  {journal} {\bibinfo
   {journal} {Phys. Rev. B}\ }\textbf {\bibinfo {volume} {100}},\ \bibinfo
  {pages} {205131} (\bibinfo {year} {2019})}\BibitemShut {NoStop}%
\bibitem [{\citenamefont {Natori}\ and\ \citenamefont
  {Knolle}(2020)}]{Natori2020}%
  \BibitemOpen
  \bibfield  {author} {\bibinfo {author} {\bibfnamefont {W.~M.~H.}\
  \bibnamefont {Natori}}\ and\ \bibinfo {author} {\bibfnamefont
  {J.}~\bibnamefont {Knolle}},\ }\bibfield  {title} {\bibinfo {title} {Dynamics
  of a two-dimensional quantum spin-orbital liquid: {Spectroscopic} signatures
  of fermionic magnons},\ }\href
  {https://doi.org/10.1103/PhysRevLett.125.067201} {\bibfield  {journal}
  {\bibinfo  {journal} {Phys. Rev. Lett.}\ }\textbf {\bibinfo {volume} {125}},\
  \bibinfo {pages} {067201} (\bibinfo {year} {2020})}\BibitemShut {NoStop}%
\bibitem [{\citenamefont {Zhou}\ \emph {et~al.}(2011)\citenamefont {Zhou},
  \citenamefont {Choi}, \citenamefont {Li}, \citenamefont {Balicas},
  \citenamefont {Wiebe}, \citenamefont {Qiu}, \citenamefont {Copley},\ and\
  \citenamefont {Gardner}}]{Zhou2011}%
  \BibitemOpen
  \bibfield  {author} {\bibinfo {author} {\bibfnamefont {H.~D.}\ \bibnamefont
  {Zhou}}, \bibinfo {author} {\bibfnamefont {E.~S.}\ \bibnamefont {Choi}},
  \bibinfo {author} {\bibfnamefont {G.}~\bibnamefont {Li}}, \bibinfo {author}
  {\bibfnamefont {L.}~\bibnamefont {Balicas}}, \bibinfo {author} {\bibfnamefont
  {C.~R.}\ \bibnamefont {Wiebe}}, \bibinfo {author} {\bibfnamefont
  {Y.}~\bibnamefont {Qiu}}, \bibinfo {author} {\bibfnamefont {J.~R.~D.}\
  \bibnamefont {Copley}},\ and\ \bibinfo {author} {\bibfnamefont {J.~S.}\
  \bibnamefont {Gardner}},\ }\bibfield  {title} {\bibinfo {title} {{Spin Liquid
  State in the $S=1/2$ Triangular Lattice
  ${\mathrm{Ba}}_{3}{\mathrm{CuSb}}_{2}{\mathrm{O}}_{9}$}},\ }\href
  {https://doi.org/10.1103/PhysRevLett.106.147204} {\bibfield  {journal}
  {\bibinfo  {journal} {Phys. Rev. Lett.}\ }\textbf {\bibinfo {volume} {106}},\
  \bibinfo {pages} {147204} (\bibinfo {year} {2011})}\BibitemShut {NoStop}%
\bibitem [{\citenamefont {Nakatsuji}\ \emph {et~al.}(2012)\citenamefont
  {Nakatsuji}, \citenamefont {Kuga}, \citenamefont {Kimura}, \citenamefont
  {Satake}, \citenamefont {Katayama}, \citenamefont {Nishibori}, \citenamefont
  {Sawa}, \citenamefont {Ishii}, \citenamefont {Hagiwara}, \citenamefont
  {Bridges} \emph {et~al.}}]{Nakatsuji2012}%
  \BibitemOpen
  \bibfield  {author} {\bibinfo {author} {\bibfnamefont {S.}~\bibnamefont
  {Nakatsuji}}, \bibinfo {author} {\bibfnamefont {K.}~\bibnamefont {Kuga}},
  \bibinfo {author} {\bibfnamefont {K.}~\bibnamefont {Kimura}}, \bibinfo
  {author} {\bibfnamefont {R.}~\bibnamefont {Satake}}, \bibinfo {author}
  {\bibfnamefont {N.}~\bibnamefont {Katayama}}, \bibinfo {author}
  {\bibfnamefont {E.}~\bibnamefont {Nishibori}}, \bibinfo {author}
  {\bibfnamefont {H.}~\bibnamefont {Sawa}}, \bibinfo {author} {\bibfnamefont
  {R.}~\bibnamefont {Ishii}}, \bibinfo {author} {\bibfnamefont
  {M.}~\bibnamefont {Hagiwara}}, \bibinfo {author} {\bibfnamefont
  {F.}~\bibnamefont {Bridges}}, \emph {et~al.},\ }\bibfield  {title} {\bibinfo
  {title} {Spin-orbital short-range order on a honeycomb-based lattice},\
  }\href {https://doi.org/10.1126/science.1212154} {\bibfield  {journal}
  {\bibinfo  {journal} {Science}\ }\textbf {\bibinfo {volume} {336}},\ \bibinfo
  {pages} {559} (\bibinfo {year} {2012})}\BibitemShut {NoStop}%
\bibitem [{\citenamefont {Quilliam}\ \emph {et~al.}(2012)\citenamefont
  {Quilliam}, \citenamefont {Bert}, \citenamefont {Kermarrec}, \citenamefont
  {Payen}, \citenamefont {Guillot-Deudon}, \citenamefont {Bonville},
  \citenamefont {Baines}, \citenamefont {Luetkens},\ and\ \citenamefont
  {Mendels}}]{Quilliam2012}%
  \BibitemOpen
  \bibfield  {author} {\bibinfo {author} {\bibfnamefont {J.~A.}\ \bibnamefont
  {Quilliam}}, \bibinfo {author} {\bibfnamefont {F.}~\bibnamefont {Bert}},
  \bibinfo {author} {\bibfnamefont {E.}~\bibnamefont {Kermarrec}}, \bibinfo
  {author} {\bibfnamefont {C.}~\bibnamefont {Payen}}, \bibinfo {author}
  {\bibfnamefont {C.}~\bibnamefont {Guillot-Deudon}}, \bibinfo {author}
  {\bibfnamefont {P.}~\bibnamefont {Bonville}}, \bibinfo {author}
  {\bibfnamefont {C.}~\bibnamefont {Baines}}, \bibinfo {author} {\bibfnamefont
  {H.}~\bibnamefont {Luetkens}},\ and\ \bibinfo {author} {\bibfnamefont
  {P.}~\bibnamefont {Mendels}},\ }\bibfield  {title} {\bibinfo {title}
  {{Singlet Ground State of the Quantum Antiferromagnet
  ${\mathrm{Ba}}_{3}{\mathrm{CuSb}}_{2}{\mathrm{O}}_{9}$}},\ }\href
  {https://doi.org/10.1103/PhysRevLett.109.117203} {\bibfield  {journal}
  {\bibinfo  {journal} {Phys. Rev. Lett.}\ }\textbf {\bibinfo {volume} {109}},\
  \bibinfo {pages} {117203} (\bibinfo {year} {2012})}\BibitemShut {NoStop}%
\bibitem [{\citenamefont {Ishiguro}\ \emph {et~al.}(2013)\citenamefont
  {Ishiguro}, \citenamefont {Kimura}, \citenamefont {Nakatsuji}, \citenamefont
  {Tsutsui}, \citenamefont {Baron}, \citenamefont {Kimura},\ and\ \citenamefont
  {Wakabayashi}}]{ishiguro2013}%
  \BibitemOpen
  \bibfield  {author} {\bibinfo {author} {\bibfnamefont {Y.}~\bibnamefont
  {Ishiguro}}, \bibinfo {author} {\bibfnamefont {K.}~\bibnamefont {Kimura}},
  \bibinfo {author} {\bibfnamefont {S.}~\bibnamefont {Nakatsuji}}, \bibinfo
  {author} {\bibfnamefont {S.}~\bibnamefont {Tsutsui}}, \bibinfo {author}
  {\bibfnamefont {A.~Q.}\ \bibnamefont {Baron}}, \bibinfo {author}
  {\bibfnamefont {T.}~\bibnamefont {Kimura}},\ and\ \bibinfo {author}
  {\bibfnamefont {Y.}~\bibnamefont {Wakabayashi}},\ }\bibfield  {title}
  {\bibinfo {title} {Dynamical spin--orbital correlation in the frustrated
  magnet {$\text{Ba}_3\text{Cu}\text{Sb}_2\text{O}_9$}},\ }\href
  {https://doi.org/10.1038/ncomms3022} {\bibfield  {journal} {\bibinfo
  {journal} {Nature communications}\ }\textbf {\bibinfo {volume} {4}},\
  \bibinfo {pages} {1} (\bibinfo {year} {2013})}\BibitemShut {NoStop}%
\bibitem [{\citenamefont {Smerald}\ and\ \citenamefont
  {Mila}(2014)}]{Smerald2014}%
  \BibitemOpen
  \bibfield  {author} {\bibinfo {author} {\bibfnamefont {A.}~\bibnamefont
  {Smerald}}\ and\ \bibinfo {author} {\bibfnamefont {F.}~\bibnamefont {Mila}},\
  }\bibfield  {title} {\bibinfo {title} {Exploring the spin-orbital ground
  state of {${\mathrm{Ba}}_{3}{\mathrm{CuSb}}_{2}{\mathrm{O}}_{9}$}},\ }\href
  {https://doi.org/10.1103/PhysRevB.90.094422} {\bibfield  {journal} {\bibinfo
  {journal} {Phys. Rev. B}\ }\textbf {\bibinfo {volume} {90}},\ \bibinfo
  {pages} {094422} (\bibinfo {year} {2014})}\BibitemShut {NoStop}%
\bibitem [{\citenamefont {Katayama}\ \emph {et~al.}(2015)\citenamefont
  {Katayama}, \citenamefont {Kimura}, \citenamefont {Han}, \citenamefont
  {Nasu}, \citenamefont {Drichko}, \citenamefont {Nakanishi}, \citenamefont
  {Halim}, \citenamefont {Ishiguro}, \citenamefont {Satake}, \citenamefont
  {Nishibori}, \citenamefont {Yoshizawa}, \citenamefont {Nakano}, \citenamefont
  {Nozue}, \citenamefont {Wakabayashi}, \citenamefont {Ishihara}, \citenamefont
  {Hagiwara}, \citenamefont {Sawa},\ and\ \citenamefont
  {Nakatsuji}}]{Katayama2015}%
  \BibitemOpen
  \bibfield  {author} {\bibinfo {author} {\bibfnamefont {N.}~\bibnamefont
  {Katayama}}, \bibinfo {author} {\bibfnamefont {K.}~\bibnamefont {Kimura}},
  \bibinfo {author} {\bibfnamefont {Y.}~\bibnamefont {Han}}, \bibinfo {author}
  {\bibfnamefont {J.}~\bibnamefont {Nasu}}, \bibinfo {author} {\bibfnamefont
  {N.}~\bibnamefont {Drichko}}, \bibinfo {author} {\bibfnamefont
  {Y.}~\bibnamefont {Nakanishi}}, \bibinfo {author} {\bibfnamefont
  {M.}~\bibnamefont {Halim}}, \bibinfo {author} {\bibfnamefont
  {Y.}~\bibnamefont {Ishiguro}}, \bibinfo {author} {\bibfnamefont
  {R.}~\bibnamefont {Satake}}, \bibinfo {author} {\bibfnamefont
  {E.}~\bibnamefont {Nishibori}}, \bibinfo {author} {\bibfnamefont
  {M.}~\bibnamefont {Yoshizawa}}, \bibinfo {author} {\bibfnamefont
  {T.}~\bibnamefont {Nakano}}, \bibinfo {author} {\bibfnamefont
  {Y.}~\bibnamefont {Nozue}}, \bibinfo {author} {\bibfnamefont
  {Y.}~\bibnamefont {Wakabayashi}}, \bibinfo {author} {\bibfnamefont
  {S.}~\bibnamefont {Ishihara}}, \bibinfo {author} {\bibfnamefont
  {M.}~\bibnamefont {Hagiwara}}, \bibinfo {author} {\bibfnamefont
  {H.}~\bibnamefont {Sawa}},\ and\ \bibinfo {author} {\bibfnamefont
  {S.}~\bibnamefont {Nakatsuji}},\ }\bibfield  {title} {\bibinfo {title}
  {{Absence of Jahn-Teller transition in the hexagonal
  ${\mathrm{Ba}}_{3}{\mathrm{CuSb}}_{2}{\mathrm{O}}_{9}$ single crystal}},\
  }\href {https://doi.org/10.1073/pnas.1508941112} {\bibfield  {journal}
  {\bibinfo  {journal} {Proceedings of the National Academy of Sciences}\
  }\textbf {\bibinfo {volume} {112}},\ \bibinfo {pages} {9305} (\bibinfo {year}
  {2015})}\BibitemShut {NoStop}%
\bibitem [{\citenamefont {Smerald}\ and\ \citenamefont
  {Mila}(2015)}]{Smerald2015}%
  \BibitemOpen
  \bibfield  {author} {\bibinfo {author} {\bibfnamefont {A.}~\bibnamefont
  {Smerald}}\ and\ \bibinfo {author} {\bibfnamefont {F.}~\bibnamefont {Mila}},\
  }\bibfield  {title} {\bibinfo {title} {Disorder-driven spin-orbital liquid
  behavior in the {${\mathrm{Ba}}_{3}X{\mathrm{Sb}}_{2}{\mathrm{O}}_{9}$}
  materials},\ }\href {https://doi.org/10.1103/PhysRevLett.115.147202}
  {\bibfield  {journal} {\bibinfo  {journal} {Phys. Rev. Lett.}\ }\textbf
  {\bibinfo {volume} {115}},\ \bibinfo {pages} {147202} (\bibinfo {year}
  {2015})}\BibitemShut {NoStop}%
\bibitem [{\citenamefont {Chen}\ \emph {et~al.}(2010)\citenamefont {Chen},
  \citenamefont {Pereira},\ and\ \citenamefont {Balents}}]{Chen2010}%
  \BibitemOpen
  \bibfield  {author} {\bibinfo {author} {\bibfnamefont {G.}~\bibnamefont
  {Chen}}, \bibinfo {author} {\bibfnamefont {R.}~\bibnamefont {Pereira}},\ and\
  \bibinfo {author} {\bibfnamefont {L.}~\bibnamefont {Balents}},\ }\bibfield
  {title} {\bibinfo {title} {Exotic phases induced by strong spin-orbit
  coupling in ordered double perovskites},\ }\href
  {https://doi.org/10.1103/PhysRevB.82.174440} {\bibfield  {journal} {\bibinfo
  {journal} {Phys. Rev. B}\ }\textbf {\bibinfo {volume} {82}},\ \bibinfo
  {pages} {174440} (\bibinfo {year} {2010})}\BibitemShut {NoStop}%
\bibitem [{\citenamefont {Yamada}\ \emph {et~al.}(2018)\citenamefont {Yamada},
  \citenamefont {Oshikawa},\ and\ \citenamefont {Jackeli}}]{Yamada2018}%
  \BibitemOpen
  \bibfield  {author} {\bibinfo {author} {\bibfnamefont {M.~G.}\ \bibnamefont
  {Yamada}}, \bibinfo {author} {\bibfnamefont {M.}~\bibnamefont {Oshikawa}},\
  and\ \bibinfo {author} {\bibfnamefont {G.}~\bibnamefont {Jackeli}},\
  }\bibfield  {title} {\bibinfo {title} {Emergent {$\mathrm{SU}(4)$} symmetry
  in {$\ensuremath{\alpha}\text{\ensuremath{-}}{\mathrm{ZrCl}}_{3}$} and
  crystalline spin-orbital liquids},\ }\href
  {https://doi.org/10.1103/PhysRevLett.121.097201} {\bibfield  {journal}
  {\bibinfo  {journal} {Phys. Rev. Lett.}\ }\textbf {\bibinfo {volume} {121}},\
  \bibinfo {pages} {097201} (\bibinfo {year} {2018})}\BibitemShut {NoStop}%
\bibitem [{\citenamefont {Ushakov}\ \emph {et~al.}(2020)\citenamefont
  {Ushakov}, \citenamefont {Solovyev},\ and\ \citenamefont
  {Streltsov}}]{Ushakov2020}%
  \BibitemOpen
  \bibfield  {author} {\bibinfo {author} {\bibfnamefont {A.~V.}\ \bibnamefont
  {Ushakov}}, \bibinfo {author} {\bibfnamefont {I.}~\bibnamefont {Solovyev}},\
  and\ \bibinfo {author} {\bibfnamefont {S.}~\bibnamefont {Streltsov}},\
  }\bibfield  {title} {\bibinfo {title} {Can the highly symmetric {SU(4)}
  spin-orbital model be realized in {$\alpha-{\mathrm{ZrCl}}_{3}$}?},\ }\href
  {https://doi.org/10.1134/S002136402022004X} {\bibfield  {journal} {\bibinfo
  {journal} {JETP Letters}\ ,\ \bibinfo {pages} {1}} (\bibinfo {year}
  {2020})}\BibitemShut {NoStop}%
\bibitem [{\citenamefont {Venderbos}\ and\ \citenamefont
  {Fernandes}(2018)}]{Venderbos2018}%
  \BibitemOpen
  \bibfield  {author} {\bibinfo {author} {\bibfnamefont {J.~W.~F.}\
  \bibnamefont {Venderbos}}\ and\ \bibinfo {author} {\bibfnamefont {R.~M.}\
  \bibnamefont {Fernandes}},\ }\bibfield  {title} {\bibinfo {title}
  {Correlations and electronic order in a two-orbital honeycomb lattice model
  for twisted bilayer graphene},\ }\href
  {https://doi.org/10.1103/PhysRevB.98.245103} {\bibfield  {journal} {\bibinfo
  {journal} {Phys. Rev. B}\ }\textbf {\bibinfo {volume} {98}},\ \bibinfo
  {pages} {245103} (\bibinfo {year} {2018})}\BibitemShut {NoStop}%
\bibitem [{\citenamefont {Aasen}\ \emph {et~al.}(2020)\citenamefont {Aasen},
  \citenamefont {Mong}, \citenamefont {Hunt}, \citenamefont {Mandrus},\ and\
  \citenamefont {Alicea}}]{Aasen2020}%
  \BibitemOpen
  \bibfield  {author} {\bibinfo {author} {\bibfnamefont {D.}~\bibnamefont
  {Aasen}}, \bibinfo {author} {\bibfnamefont {R.~S.~K.}\ \bibnamefont {Mong}},
  \bibinfo {author} {\bibfnamefont {B.~M.}\ \bibnamefont {Hunt}}, \bibinfo
  {author} {\bibfnamefont {D.}~\bibnamefont {Mandrus}},\ and\ \bibinfo {author}
  {\bibfnamefont {J.}~\bibnamefont {Alicea}},\ }\bibfield  {title} {\bibinfo
  {title} {Electrical probes of the non-abelian spin liquid in {Kitaev}
  materials},\ }\href {https://doi.org/10.1103/PhysRevX.10.031014} {\bibfield
  {journal} {\bibinfo  {journal} {Phys. Rev. X}\ }\textbf {\bibinfo {volume}
  {10}},\ \bibinfo {pages} {031014} (\bibinfo {year} {2020})}\BibitemShut
  {NoStop}%
\bibitem [{\citenamefont {K\"onig}\ \emph {et~al.}(2020)\citenamefont
  {K\"onig}, \citenamefont {Randeria},\ and\ \citenamefont
  {J\"ack}}]{Konig2020}%
  \BibitemOpen
  \bibfield  {author} {\bibinfo {author} {\bibfnamefont {E.~J.}\ \bibnamefont
  {K\"onig}}, \bibinfo {author} {\bibfnamefont {M.~T.}\ \bibnamefont
  {Randeria}},\ and\ \bibinfo {author} {\bibfnamefont {B.}~\bibnamefont
  {J\"ack}},\ }\bibfield  {title} {\bibinfo {title} {Tunneling spectroscopy of
  quantum spin liquids},\ }\href
  {https://doi.org/10.1103/PhysRevLett.125.267206} {\bibfield  {journal}
  {\bibinfo  {journal} {Phys. Rev. Lett.}\ }\textbf {\bibinfo {volume} {125}},\
  \bibinfo {pages} {267206} (\bibinfo {year} {2020})}\BibitemShut {NoStop}%
\bibitem [{\citenamefont {Kasahara}\ \emph {et~al.}(2018)\citenamefont
  {Kasahara}, \citenamefont {Ohnishi}, \citenamefont {Mizukami}, \citenamefont
  {Tanaka}, \citenamefont {Ma}, \citenamefont {Sugii}, \citenamefont {Kurita},
  \citenamefont {Tanaka}, \citenamefont {Nasu}, \citenamefont {Motome} \emph
  {et~al.}}]{kasahara2018}%
  \BibitemOpen
  \bibfield  {author} {\bibinfo {author} {\bibfnamefont {Y.}~\bibnamefont
  {Kasahara}}, \bibinfo {author} {\bibfnamefont {T.}~\bibnamefont {Ohnishi}},
  \bibinfo {author} {\bibfnamefont {Y.}~\bibnamefont {Mizukami}}, \bibinfo
  {author} {\bibfnamefont {O.}~\bibnamefont {Tanaka}}, \bibinfo {author}
  {\bibfnamefont {S.}~\bibnamefont {Ma}}, \bibinfo {author} {\bibfnamefont
  {K.}~\bibnamefont {Sugii}}, \bibinfo {author} {\bibfnamefont
  {N.}~\bibnamefont {Kurita}}, \bibinfo {author} {\bibfnamefont
  {H.}~\bibnamefont {Tanaka}}, \bibinfo {author} {\bibfnamefont
  {J.}~\bibnamefont {Nasu}}, \bibinfo {author} {\bibfnamefont {Y.}~\bibnamefont
  {Motome}}, \emph {et~al.},\ }\bibfield  {title} {\bibinfo {title} {{Majorana
  quantization and half-integer thermal quantum Hall effect in a Kitaev spin
  liquid}},\ }\href {https://doi.org/10.1038/s41586-018-0274-0} {\bibfield
  {journal} {\bibinfo  {journal} {Nature}\ }\textbf {\bibinfo {volume} {559}},\
  \bibinfo {pages} {227} (\bibinfo {year} {2018})}\BibitemShut {NoStop}%
\bibitem [{\citenamefont {Hirobe}\ \emph {et~al.}(2017)\citenamefont {Hirobe},
  \citenamefont {Sato}, \citenamefont {Kawamata}, \citenamefont {Shiomi},
  \citenamefont {Uchida}, \citenamefont {Iguchi}, \citenamefont {Koike},
  \citenamefont {Maekawa},\ and\ \citenamefont {Saitoh}}]{Hirobe2017}%
  \BibitemOpen
  \bibfield  {author} {\bibinfo {author} {\bibfnamefont {D.}~\bibnamefont
  {Hirobe}}, \bibinfo {author} {\bibfnamefont {M.}~\bibnamefont {Sato}},
  \bibinfo {author} {\bibfnamefont {T.}~\bibnamefont {Kawamata}}, \bibinfo
  {author} {\bibfnamefont {Y.}~\bibnamefont {Shiomi}}, \bibinfo {author}
  {\bibfnamefont {K.-i.}\ \bibnamefont {Uchida}}, \bibinfo {author}
  {\bibfnamefont {R.}~\bibnamefont {Iguchi}}, \bibinfo {author} {\bibfnamefont
  {Y.}~\bibnamefont {Koike}}, \bibinfo {author} {\bibfnamefont
  {S.}~\bibnamefont {Maekawa}},\ and\ \bibinfo {author} {\bibfnamefont
  {E.}~\bibnamefont {Saitoh}},\ }\bibfield  {title} {\bibinfo {title}
  {One-dimensional spinon spin currents},\ }\href
  {https://doi.org/10.1038/nphys3895} {\bibfield  {journal} {\bibinfo
  {journal} {Nature Physics}\ }\textbf {\bibinfo {volume} {13}},\ \bibinfo
  {pages} {30} (\bibinfo {year} {2017})}\BibitemShut {NoStop}%
\bibitem [{\citenamefont {Hirobe}\ \emph {et~al.}(2018)\citenamefont {Hirobe},
  \citenamefont {Kawamata}, \citenamefont {Oyanagi}, \citenamefont {Koike},\
  and\ \citenamefont {Saitoh}}]{Hirobe2018}%
  \BibitemOpen
  \bibfield  {author} {\bibinfo {author} {\bibfnamefont {D.}~\bibnamefont
  {Hirobe}}, \bibinfo {author} {\bibfnamefont {T.}~\bibnamefont {Kawamata}},
  \bibinfo {author} {\bibfnamefont {K.}~\bibnamefont {Oyanagi}}, \bibinfo
  {author} {\bibfnamefont {Y.}~\bibnamefont {Koike}},\ and\ \bibinfo {author}
  {\bibfnamefont {E.}~\bibnamefont {Saitoh}},\ }\bibfield  {title} {\bibinfo
  {title} {Generation of spin currents from one-dimensional quantum spin
  liquid},\ }\href {https://doi.org/10.1063/1.5021022} {\bibfield  {journal}
  {\bibinfo  {journal} {Journal of Applied Physics}\ }\textbf {\bibinfo
  {volume} {123}},\ \bibinfo {pages} {123903} (\bibinfo {year}
  {2018})}\BibitemShut {NoStop}%
\bibitem [{\citenamefont {Koga}\ \emph {et~al.}(2020)\citenamefont {Koga},
  \citenamefont {Minakawa}, \citenamefont {Murakami},\ and\ \citenamefont
  {Nasu}}]{Koga2020}%
  \BibitemOpen
  \bibfield  {author} {\bibinfo {author} {\bibfnamefont {A.}~\bibnamefont
  {Koga}}, \bibinfo {author} {\bibfnamefont {T.}~\bibnamefont {Minakawa}},
  \bibinfo {author} {\bibfnamefont {Y.}~\bibnamefont {Murakami}},\ and\
  \bibinfo {author} {\bibfnamefont {J.}~\bibnamefont {Nasu}},\ }\bibfield
  {title} {\bibinfo {title} {Spin transport in the quantum spin liquid state in
  the s = 1 {Kitaev} model: {Role} of the fractionalized quasiparticles},\
  }\href {https://doi.org/10.7566/JPSJ.89.033701} {\bibfield  {journal}
  {\bibinfo  {journal} {Journal of the Physical Society of Japan}\ }\textbf
  {\bibinfo {volume} {89}},\ \bibinfo {pages} {033701} (\bibinfo {year}
  {2020})}\BibitemShut {NoStop}%
\bibitem [{\citenamefont {Minakawa}\ \emph {et~al.}(2020)\citenamefont
  {Minakawa}, \citenamefont {Murakami}, \citenamefont {Koga},\ and\
  \citenamefont {Nasu}}]{Minakawa2020}%
  \BibitemOpen
  \bibfield  {author} {\bibinfo {author} {\bibfnamefont {T.}~\bibnamefont
  {Minakawa}}, \bibinfo {author} {\bibfnamefont {Y.}~\bibnamefont {Murakami}},
  \bibinfo {author} {\bibfnamefont {A.}~\bibnamefont {Koga}},\ and\ \bibinfo
  {author} {\bibfnamefont {J.}~\bibnamefont {Nasu}},\ }\bibfield  {title}
  {\bibinfo {title} {Majorana-mediated spin transport in {Kitaev} quantum spin
  liquids},\ }\href {https://doi.org/10.1103/PhysRevLett.125.047204} {\bibfield
   {journal} {\bibinfo  {journal} {Phys. Rev. Lett.}\ }\textbf {\bibinfo
  {volume} {125}},\ \bibinfo {pages} {047204} (\bibinfo {year}
  {2020})}\BibitemShut {NoStop}%
\bibitem [{\citenamefont {Mizoguchi}\ \emph {et~al.}(2020)\citenamefont
  {Mizoguchi}, \citenamefont {Koma},\ and\ \citenamefont
  {Yoshida}}]{Mizoguchi2020}%
  \BibitemOpen
  \bibfield  {author} {\bibinfo {author} {\bibfnamefont {T.}~\bibnamefont
  {Mizoguchi}}, \bibinfo {author} {\bibfnamefont {T.}~\bibnamefont {Koma}},\
  and\ \bibinfo {author} {\bibfnamefont {Y.}~\bibnamefont {Yoshida}},\
  }\bibfield  {title} {\bibinfo {title} {Oriented propagation of magnetization
  due to chiral edge modes in {Kitaev}-type models},\ }\href
  {https://doi.org/10.1103/PhysRevB.101.014442} {\bibfield  {journal} {\bibinfo
   {journal} {Phys. Rev. B}\ }\textbf {\bibinfo {volume} {101}},\ \bibinfo
  {pages} {014442} (\bibinfo {year} {2020})}\BibitemShut {NoStop}%
\bibitem [{\citenamefont {Han}\ \emph {et~al.}(2020)\citenamefont {Han},
  \citenamefont {Maekawa},\ and\ \citenamefont {Xie}}]{han2020}%
  \BibitemOpen
  \bibfield  {author} {\bibinfo {author} {\bibfnamefont {W.}~\bibnamefont
  {Han}}, \bibinfo {author} {\bibfnamefont {S.}~\bibnamefont {Maekawa}},\ and\
  \bibinfo {author} {\bibfnamefont {X.-C.}\ \bibnamefont {Xie}},\ }\bibfield
  {title} {\bibinfo {title} {Spin current as a probe of quantum materials},\
  }\href {https://doi.org/10.1038/s41563-019-0456-7} {\bibfield  {journal}
  {\bibinfo  {journal} {Nature materials}\ }\textbf {\bibinfo {volume} {19}},\
  \bibinfo {pages} {139} (\bibinfo {year} {2020})}\BibitemShut {NoStop}%
\bibitem [{\citenamefont {Chen}\ \emph {et~al.}(2013)\citenamefont {Chen},
  \citenamefont {Sun}, \citenamefont {Wang},\ and\ \citenamefont
  {Xie}}]{Chen2013}%
  \BibitemOpen
  \bibfield  {author} {\bibinfo {author} {\bibfnamefont {C.-Z.}\ \bibnamefont
  {Chen}}, \bibinfo {author} {\bibfnamefont {Q.-f.}\ \bibnamefont {Sun}},
  \bibinfo {author} {\bibfnamefont {F.}~\bibnamefont {Wang}},\ and\ \bibinfo
  {author} {\bibfnamefont {X.~C.}\ \bibnamefont {Xie}},\ }\bibfield  {title}
  {\bibinfo {title} {Detection of spinons via spin transport},\ }\href
  {https://doi.org/10.1103/PhysRevB.88.041405} {\bibfield  {journal} {\bibinfo
  {journal} {Phys. Rev. B}\ }\textbf {\bibinfo {volume} {88}},\ \bibinfo
  {pages} {041405(R)} (\bibinfo {year} {2013})}\BibitemShut {NoStop}%
\bibitem [{\citenamefont {Chatterjee}\ and\ \citenamefont
  {Sachdev}(2015)}]{Chatterjee2015}%
  \BibitemOpen
  \bibfield  {author} {\bibinfo {author} {\bibfnamefont {S.}~\bibnamefont
  {Chatterjee}}\ and\ \bibinfo {author} {\bibfnamefont {S.}~\bibnamefont
  {Sachdev}},\ }\bibfield  {title} {\bibinfo {title} {Probing excitations in
  insulators via injection of spin currents},\ }\href
  {https://doi.org/10.1103/PhysRevB.92.165113} {\bibfield  {journal} {\bibinfo
  {journal} {Phys. Rev. B}\ }\textbf {\bibinfo {volume} {92}},\ \bibinfo
  {pages} {165113} (\bibinfo {year} {2015})}\BibitemShut {NoStop}%
\bibitem [{\citenamefont {de~Carvalho}\ \emph {et~al.}(2018)\citenamefont
  {de~Carvalho}, \citenamefont {Freire}, \citenamefont {Miranda},\ and\
  \citenamefont {Pereira}}]{DeCarvalho2018}%
  \BibitemOpen
  \bibfield  {author} {\bibinfo {author} {\bibfnamefont {V.~S.}\ \bibnamefont
  {de~Carvalho}}, \bibinfo {author} {\bibfnamefont {H.}~\bibnamefont {Freire}},
  \bibinfo {author} {\bibfnamefont {E.}~\bibnamefont {Miranda}},\ and\ \bibinfo
  {author} {\bibfnamefont {R.~G.}\ \bibnamefont {Pereira}},\ }\bibfield
  {title} {\bibinfo {title} {{Edge magnetization and spin transport in an
  SU(2)-symmetric Kitaev spin liquid}},\ }\href
  {https://doi.org/10.1103/PhysRevB.98.155105} {\bibfield  {journal} {\bibinfo
  {journal} {Phys. Rev. B}\ }\textbf {\bibinfo {volume} {98}},\ \bibinfo
  {pages} {155105} (\bibinfo {year} {2018})}\BibitemShut {NoStop}%
\bibitem [{\citenamefont {Spencer}\ and\ \citenamefont
  {Doniach}(1967)}]{Spencer1967}%
  \BibitemOpen
  \bibfield  {author} {\bibinfo {author} {\bibfnamefont {H.~J.}\ \bibnamefont
  {Spencer}}\ and\ \bibinfo {author} {\bibfnamefont {S.}~\bibnamefont
  {Doniach}},\ }\bibfield  {title} {\bibinfo {title} {Low-temperature anomaly
  of electron-spin resonance in dilute alloys},\ }\href
  {https://doi.org/10.1103/PhysRevLett.18.994} {\bibfield  {journal} {\bibinfo
  {journal} {Phys. Rev. Lett.}\ }\textbf {\bibinfo {volume} {18}},\ \bibinfo
  {pages} {994} (\bibinfo {year} {1967})}\BibitemShut {NoStop}%
\bibitem [{\citenamefont {Coleman}\ \emph {et~al.}(1993)\citenamefont
  {Coleman}, \citenamefont {Miranda},\ and\ \citenamefont
  {Tsvelik}}]{Coleman1993}%
  \BibitemOpen
  \bibfield  {author} {\bibinfo {author} {\bibfnamefont {P.}~\bibnamefont
  {Coleman}}, \bibinfo {author} {\bibfnamefont {E.}~\bibnamefont {Miranda}},\
  and\ \bibinfo {author} {\bibfnamefont {A.}~\bibnamefont {Tsvelik}},\
  }\bibfield  {title} {\bibinfo {title} {Possible realization of odd-frequency
  pairing in heavy fermion compounds},\ }\href
  {https://doi.org/10.1103/PhysRevLett.70.2960} {\bibfield  {journal} {\bibinfo
   {journal} {Phys. Rev. Lett.}\ }\textbf {\bibinfo {volume} {70}},\ \bibinfo
  {pages} {2960} (\bibinfo {year} {1993})}\BibitemShut {NoStop}%
\bibitem [{\citenamefont {Shnirman}\ and\ \citenamefont
  {Makhlin}(2003)}]{Shnirman2003}%
  \BibitemOpen
  \bibfield  {author} {\bibinfo {author} {\bibfnamefont {A.}~\bibnamefont
  {Shnirman}}\ and\ \bibinfo {author} {\bibfnamefont {Y.}~\bibnamefont
  {Makhlin}},\ }\bibfield  {title} {\bibinfo {title} {{Spin-Spin Correlators in
  the Majorana Representation}},\ }\href
  {https://doi.org/10.1103/PhysRevLett.91.207204} {\bibfield  {journal}
  {\bibinfo  {journal} {Phys. Rev. Lett.}\ }\textbf {\bibinfo {volume} {91}},\
  \bibinfo {pages} {207204} (\bibinfo {year} {2003})}\BibitemShut {NoStop}%
\bibitem [{\citenamefont {Mao}\ \emph {et~al.}(2003)\citenamefont {Mao},
  \citenamefont {Coleman}, \citenamefont {Hooley},\ and\ \citenamefont
  {Langreth}}]{Mao2003}%
  \BibitemOpen
  \bibfield  {author} {\bibinfo {author} {\bibfnamefont {W.}~\bibnamefont
  {Mao}}, \bibinfo {author} {\bibfnamefont {P.}~\bibnamefont {Coleman}},
  \bibinfo {author} {\bibfnamefont {C.}~\bibnamefont {Hooley}},\ and\ \bibinfo
  {author} {\bibfnamefont {D.}~\bibnamefont {Langreth}},\ }\bibfield  {title}
  {\bibinfo {title} {{Spin Dynamics from Majorana Fermions}},\ }\href
  {https://doi.org/10.1103/PhysRevLett.91.207203} {\bibfield  {journal}
  {\bibinfo  {journal} {Phys. Rev. Lett.}\ }\textbf {\bibinfo {volume} {91}},\
  \bibinfo {pages} {207203} (\bibinfo {year} {2003})}\BibitemShut {NoStop}%
\bibitem [{\citenamefont {Haldane}(1988)}]{Haldane1988}%
  \BibitemOpen
  \bibfield  {author} {\bibinfo {author} {\bibfnamefont {F.~D.~M.}\
  \bibnamefont {Haldane}},\ }\bibfield  {title} {\bibinfo {title} {{Model for a
  Quantum Hall Effect without Landau Levels: Condensed-Matter Realization of
  the "Parity Anomaly"}},\ }\href {https://doi.org/10.1103/PhysRevLett.61.2015}
  {\bibfield  {journal} {\bibinfo  {journal} {Phys. Rev. Lett.}\ }\textbf
  {\bibinfo {volume} {61}},\ \bibinfo {pages} {2015} (\bibinfo {year}
  {1988})}\BibitemShut {NoStop}%
\bibitem [{\citenamefont {Cornelissen}\ \emph {et~al.}(2015)\citenamefont
  {Cornelissen}, \citenamefont {Liu}, \citenamefont {Duine}, \citenamefont
  {Youssef},\ and\ \citenamefont {Van~Wees}}]{cornelissen2015}%
  \BibitemOpen
  \bibfield  {author} {\bibinfo {author} {\bibfnamefont {L.}~\bibnamefont
  {Cornelissen}}, \bibinfo {author} {\bibfnamefont {J.}~\bibnamefont {Liu}},
  \bibinfo {author} {\bibfnamefont {R.}~\bibnamefont {Duine}}, \bibinfo
  {author} {\bibfnamefont {J.~B.}\ \bibnamefont {Youssef}},\ and\ \bibinfo
  {author} {\bibfnamefont {B.}~\bibnamefont {Van~Wees}},\ }\bibfield  {title}
  {\bibinfo {title} {Long-distance transport of magnon spin information in a
  magnetic insulator at room temperature},\ }\href
  {https://doi.org/10.1038/nphys3465} {\bibfield  {journal} {\bibinfo
  {journal} {Nature Physics}\ }\textbf {\bibinfo {volume} {11}},\ \bibinfo
  {pages} {1022} (\bibinfo {year} {2015})}\BibitemShut {NoStop}%
\bibitem [{\citenamefont {Wesenberg}\ \emph {et~al.}(2017)\citenamefont
  {Wesenberg}, \citenamefont {Liu}, \citenamefont {Balzar}, \citenamefont
  {Wu},\ and\ \citenamefont {Zink}}]{wesenberg2017}%
  \BibitemOpen
  \bibfield  {author} {\bibinfo {author} {\bibfnamefont {D.}~\bibnamefont
  {Wesenberg}}, \bibinfo {author} {\bibfnamefont {T.}~\bibnamefont {Liu}},
  \bibinfo {author} {\bibfnamefont {D.}~\bibnamefont {Balzar}}, \bibinfo
  {author} {\bibfnamefont {M.}~\bibnamefont {Wu}},\ and\ \bibinfo {author}
  {\bibfnamefont {B.~L.}\ \bibnamefont {Zink}},\ }\bibfield  {title} {\bibinfo
  {title} {Long-distance spin transport in a disordered magnetic insulator},\
  }\href {https://doi.org/10.1038/nphys4175} {\bibfield  {journal} {\bibinfo
  {journal} {Nature Physics}\ }\textbf {\bibinfo {volume} {13}},\ \bibinfo
  {pages} {987} (\bibinfo {year} {2017})}\BibitemShut {NoStop}%
\bibitem [{\citenamefont {Lebrun}\ \emph {et~al.}(2018)\citenamefont {Lebrun},
  \citenamefont {Ross}, \citenamefont {Bender}, \citenamefont {Qaiumzadeh},
  \citenamefont {Baldrati}, \citenamefont {Cramer}, \citenamefont {Brataas},
  \citenamefont {Duine},\ and\ \citenamefont {Kl{\"a}ui}}]{lebrun2018}%
  \BibitemOpen
  \bibfield  {author} {\bibinfo {author} {\bibfnamefont {R.}~\bibnamefont
  {Lebrun}}, \bibinfo {author} {\bibfnamefont {A.}~\bibnamefont {Ross}},
  \bibinfo {author} {\bibfnamefont {S.}~\bibnamefont {Bender}}, \bibinfo
  {author} {\bibfnamefont {A.}~\bibnamefont {Qaiumzadeh}}, \bibinfo {author}
  {\bibfnamefont {L.}~\bibnamefont {Baldrati}}, \bibinfo {author}
  {\bibfnamefont {J.}~\bibnamefont {Cramer}}, \bibinfo {author} {\bibfnamefont
  {A.}~\bibnamefont {Brataas}}, \bibinfo {author} {\bibfnamefont
  {R.}~\bibnamefont {Duine}},\ and\ \bibinfo {author} {\bibfnamefont
  {M.}~\bibnamefont {Kl{\"a}ui}},\ }\bibfield  {title} {\bibinfo {title}
  {Tunable long-distance spin transport in a crystalline antiferromagnetic iron
  oxide},\ }\href {https://doi.org/10.1038/s41586-018-0490-7} {\bibfield
  {journal} {\bibinfo  {journal} {Nature}\ }\textbf {\bibinfo {volume} {561}},\
  \bibinfo {pages} {222} (\bibinfo {year} {2018})}\BibitemShut {NoStop}%
\bibitem [{\citenamefont {Rammer}(2007)}]{rammer2007}%
  \BibitemOpen
  \bibfield  {author} {\bibinfo {author} {\bibfnamefont {J.}~\bibnamefont
  {Rammer}},\ }\href@noop {} {\emph {\bibinfo {title} {Quantum field theory of
  non-equilibrium states}}},\ Vol.~\bibinfo {volume} {22}\ (\bibinfo
  {publisher} {Cambridge University Press Cambridge},\ \bibinfo {year}
  {2007})\BibitemShut {NoStop}%
\bibitem [{\citenamefont {Mahan}(2013)}]{mahan2013}%
  \BibitemOpen
  \bibfield  {author} {\bibinfo {author} {\bibfnamefont {G.~D.}\ \bibnamefont
  {Mahan}},\ }\href@noop {} {\emph {\bibinfo {title} {Many-particle physics}}}\
  (\bibinfo  {publisher} {Springer Science \& Business Media},\ \bibinfo {year}
  {2013})\BibitemShut {NoStop}%
\bibitem [{\citenamefont {Caroli}\ \emph {et~al.}(1971)\citenamefont {Caroli},
  \citenamefont {Combescot}, \citenamefont {Nozieres},\ and\ \citenamefont
  {Saint-James}}]{Caroli1971}%
  \BibitemOpen
  \bibfield  {author} {\bibinfo {author} {\bibfnamefont {C.}~\bibnamefont
  {Caroli}}, \bibinfo {author} {\bibfnamefont {R.}~\bibnamefont {Combescot}},
  \bibinfo {author} {\bibfnamefont {P.}~\bibnamefont {Nozieres}},\ and\
  \bibinfo {author} {\bibfnamefont {D.}~\bibnamefont {Saint-James}},\
  }\bibfield  {title} {\bibinfo {title} {Direct calculation of the tunneling
  current},\ }\href {https://doi.org/10.1088/0022-3719/4/8/018} {\bibfield
  {journal} {\bibinfo  {journal} {Journal of Physics C: Solid State Physics}\
  }\textbf {\bibinfo {volume} {4}},\ \bibinfo {pages} {916} (\bibinfo {year}
  {1971})}\BibitemShut {NoStop}%
\bibitem [{\citenamefont {Meir}\ and\ \citenamefont
  {Wingreen}(1992)}]{Meir1992}%
  \BibitemOpen
  \bibfield  {author} {\bibinfo {author} {\bibfnamefont {Y.}~\bibnamefont
  {Meir}}\ and\ \bibinfo {author} {\bibfnamefont {N.~S.}\ \bibnamefont
  {Wingreen}},\ }\bibfield  {title} {\bibinfo {title} {Landauer formula for the
  current through an interacting electron region},\ }\href
  {https://doi.org/10.1103/PhysRevLett.68.2512} {\bibfield  {journal} {\bibinfo
   {journal} {Phys. Rev. Lett.}\ }\textbf {\bibinfo {volume} {68}},\ \bibinfo
  {pages} {2512} (\bibinfo {year} {1992})}\BibitemShut {NoStop}%
\bibitem [{\citenamefont {Datta}(1997)}]{Datta1997}%
  \BibitemOpen
  \bibfield  {author} {\bibinfo {author} {\bibfnamefont {S.}~\bibnamefont
  {Datta}},\ }\href@noop {} {\emph {\bibinfo {title} {Electronic transport in
  mesoscopic systems}}}\ (\bibinfo  {publisher} {Cambridge university press},\
  \bibinfo {year} {1997})\BibitemShut {NoStop}%
\bibitem [{\citenamefont {Zhuang}\ \emph {et~al.}(2020)\citenamefont {Zhuang},
  \citenamefont {Merino},\ and\ \citenamefont {Marston}}]{Zhuang2020}%
  \BibitemOpen
  \bibfield  {author} {\bibinfo {author} {\bibfnamefont {Z.}~\bibnamefont
  {Zhuang}}, \bibinfo {author} {\bibfnamefont {J.}~\bibnamefont {Merino}},\
  and\ \bibinfo {author} {\bibfnamefont {J.~B.}\ \bibnamefont {Marston}},\
  }\bibfield  {title} {\bibinfo {title} {{Transport in conductors and
  rectifiers: Mean-field Redfield equations and nonequilibrium Green's
  functions}},\ }\href {https://doi.org/10.1103/PhysRevB.102.125147} {\bibfield
   {journal} {\bibinfo  {journal} {Phys. Rev. B}\ }\textbf {\bibinfo {volume}
  {102}},\ \bibinfo {pages} {125147} (\bibinfo {year} {2020})}\BibitemShut
  {NoStop}%
\bibitem [{\citenamefont {Lieb}(1994)}]{Lieb1994}%
  \BibitemOpen
  \bibfield  {author} {\bibinfo {author} {\bibfnamefont {E.~H.}\ \bibnamefont
  {Lieb}},\ }\bibfield  {title} {\bibinfo {title} {Flux phase of the
  half-filled band},\ }\href {https://doi.org/10.1103/PhysRevLett.73.2158}
  {\bibfield  {journal} {\bibinfo  {journal} {Phys. Rev. Lett.}\ }\textbf
  {\bibinfo {volume} {73}},\ \bibinfo {pages} {2158} (\bibinfo {year}
  {1994})}\BibitemShut {NoStop}%
\bibitem [{SM()}]{SM}%
  \BibitemOpen
  \href@noop {} {\bibinfo {title} {See supplemental material for more
  details.}}\BibitemShut {Stop}%
\bibitem [{\citenamefont {Kohmoto}\ and\ \citenamefont
  {Hasegawa}(2007)}]{Kohmoto2007}%
  \BibitemOpen
  \bibfield  {author} {\bibinfo {author} {\bibfnamefont {M.}~\bibnamefont
  {Kohmoto}}\ and\ \bibinfo {author} {\bibfnamefont {Y.}~\bibnamefont
  {Hasegawa}},\ }\bibfield  {title} {\bibinfo {title} {Zero modes and edge
  states of the honeycomb lattice},\ }\href
  {https://doi.org/10.1103/PhysRevB.76.205402} {\bibfield  {journal} {\bibinfo
  {journal} {Phys. Rev. B}\ }\textbf {\bibinfo {volume} {76}},\ \bibinfo
  {pages} {205402} (\bibinfo {year} {2007})}\BibitemShut {NoStop}%
\bibitem [{\citenamefont {Thakurathi}\ \emph {et~al.}(2014)\citenamefont
  {Thakurathi}, \citenamefont {Sengupta},\ and\ \citenamefont
  {Sen}}]{Thakurathi2014}%
  \BibitemOpen
  \bibfield  {author} {\bibinfo {author} {\bibfnamefont {M.}~\bibnamefont
  {Thakurathi}}, \bibinfo {author} {\bibfnamefont {K.}~\bibnamefont
  {Sengupta}},\ and\ \bibinfo {author} {\bibfnamefont {D.}~\bibnamefont
  {Sen}},\ }\bibfield  {title} {\bibinfo {title} {Majorana edge modes in the
  {Kitaev} model},\ }\href {https://doi.org/10.1103/PhysRevB.89.235434}
  {\bibfield  {journal} {\bibinfo  {journal} {Phys. Rev. B}\ }\textbf {\bibinfo
  {volume} {89}},\ \bibinfo {pages} {235434} (\bibinfo {year}
  {2014})}\BibitemShut {NoStop}%
\bibitem [{\citenamefont {Mizoguchi}\ and\ \citenamefont
  {Koma}(2019)}]{Mizoguchi2019}%
  \BibitemOpen
  \bibfield  {author} {\bibinfo {author} {\bibfnamefont {T.}~\bibnamefont
  {Mizoguchi}}\ and\ \bibinfo {author} {\bibfnamefont {T.}~\bibnamefont
  {Koma}},\ }\bibfield  {title} {\bibinfo {title} {Majorana edge magnetization
  in the {Kitaev} honeycomb model},\ }\href
  {https://doi.org/10.1103/PhysRevB.99.184418} {\bibfield  {journal} {\bibinfo
  {journal} {Phys. Rev. B}\ }\textbf {\bibinfo {volume} {99}},\ \bibinfo
  {pages} {184418} (\bibinfo {year} {2019})}\BibitemShut {NoStop}%
\bibitem [{\citenamefont {Castro~Neto}\ \emph {et~al.}(2009)\citenamefont
  {Castro~Neto}, \citenamefont {Guinea}, \citenamefont {Peres}, \citenamefont
  {Novoselov},\ and\ \citenamefont {Geim}}]{CastroNeto2009}%
  \BibitemOpen
  \bibfield  {author} {\bibinfo {author} {\bibfnamefont {A.~H.}\ \bibnamefont
  {Castro~Neto}}, \bibinfo {author} {\bibfnamefont {F.}~\bibnamefont {Guinea}},
  \bibinfo {author} {\bibfnamefont {N.~M.~R.}\ \bibnamefont {Peres}}, \bibinfo
  {author} {\bibfnamefont {K.~S.}\ \bibnamefont {Novoselov}},\ and\ \bibinfo
  {author} {\bibfnamefont {A.~K.}\ \bibnamefont {Geim}},\ }\bibfield  {title}
  {\bibinfo {title} {The electronic properties of graphene},\ }\href
  {https://doi.org/10.1103/RevModPhys.81.109} {\bibfield  {journal} {\bibinfo
  {journal} {Rev. Mod. Phys.}\ }\textbf {\bibinfo {volume} {81}},\ \bibinfo
  {pages} {109} (\bibinfo {year} {2009})}\BibitemShut {NoStop}%
\bibitem [{\citenamefont {Das~Sarma}\ \emph {et~al.}(2011)\citenamefont
  {Das~Sarma}, \citenamefont {Adam}, \citenamefont {Hwang},\ and\ \citenamefont
  {Rossi}}]{DasSarma2011}%
  \BibitemOpen
  \bibfield  {author} {\bibinfo {author} {\bibfnamefont {S.}~\bibnamefont
  {Das~Sarma}}, \bibinfo {author} {\bibfnamefont {S.}~\bibnamefont {Adam}},
  \bibinfo {author} {\bibfnamefont {E.~H.}\ \bibnamefont {Hwang}},\ and\
  \bibinfo {author} {\bibfnamefont {E.}~\bibnamefont {Rossi}},\ }\bibfield
  {title} {\bibinfo {title} {Electronic transport in two-dimensional
  graphene},\ }\href {https://doi.org/10.1103/RevModPhys.83.407} {\bibfield
  {journal} {\bibinfo  {journal} {Rev. Mod. Phys.}\ }\textbf {\bibinfo {volume}
  {83}},\ \bibinfo {pages} {407} (\bibinfo {year} {2011})}\BibitemShut
  {NoStop}%
\end{thebibliography}%


\begin{thebibliography}{4}%
\makeatletter
\providecommand \@ifxundefined [1]{%
 \@ifx{#1\undefined}
}%
\providecommand \@ifnum [1]{%
 \ifnum #1\expandafter \@firstoftwo
 \else \expandafter \@secondoftwo
 \fi
}%
\providecommand \@ifx [1]{%
 \ifx #1\expandafter \@firstoftwo
 \else \expandafter \@secondoftwo
 \fi
}%
\providecommand \natexlab [1]{#1}%
\providecommand \enquote  [1]{``#1''}%
\providecommand \bibnamefont  [1]{#1}%
\providecommand \bibfnamefont [1]{#1}%
\providecommand \citenamefont [1]{#1}%
\providecommand \href@noop [0]{\@secondoftwo}%
\providecommand \href [0]{\begingroup \@sanitize@url \@href}%
\providecommand \@href[1]{\@@startlink{#1}\@@href}%
\providecommand \@@href[1]{\endgroup#1\@@endlink}%
\providecommand \@sanitize@url [0]{\catcode `\\12\catcode `\$12\catcode
  `\&12\catcode `\#12\catcode `\^12\catcode `\_12\catcode `\%12\relax}%
\providecommand \@@startlink[1]{}%
\providecommand \@@endlink[0]{}%
\providecommand \url  [0]{\begingroup\@sanitize@url \@url }%
\providecommand \@url [1]{\endgroup\@href {#1}{\urlprefix }}%
\providecommand \urlprefix  [0]{URL }%
\providecommand \Eprint [0]{\href }%
\providecommand \doibase [0]{https://doi.org/}%
\providecommand \selectlanguage [0]{\@gobble}%
\providecommand \bibinfo  [0]{\@secondoftwo}%
\providecommand \bibfield  [0]{\@secondoftwo}%
\providecommand \translation [1]{[#1]}%
\providecommand \BibitemOpen [0]{}%
\providecommand \bibitemStop [0]{}%
\providecommand \bibitemNoStop [0]{.\EOS\space}%
\providecommand \EOS [0]{\spacefactor3000\relax}%
\providecommand \BibitemShut  [1]{\csname bibitem#1\endcsname}%
\let\auto@bib@innerbib\@empty
%</preamble>
\bibitem [{\citenamefont {Rammer}(2007)}]{rammer2007}%
  \BibitemOpen
  \bibfield  {author} {\bibinfo {author} {\bibfnamefont {J.}~\bibnamefont
  {Rammer}},\ }\href@noop {} {\emph {\bibinfo {title} {Quantum field theory of
  non-equilibrium states}}},\ Vol.~\bibinfo {volume} {22}\ (\bibinfo
  {publisher} {Cambridge University Press Cambridge},\ \bibinfo {year}
  {2007})\BibitemShut {NoStop}%
\bibitem [{\citenamefont {Chen}\ \emph {et~al.}(2013)\citenamefont {Chen},
  \citenamefont {Sun}, \citenamefont {Wang},\ and\ \citenamefont
  {Xie}}]{Chen2013}%
  \BibitemOpen
  \bibfield  {author} {\bibinfo {author} {\bibfnamefont {C.-Z.}\ \bibnamefont
  {Chen}}, \bibinfo {author} {\bibfnamefont {Q.-f.}\ \bibnamefont {Sun}},
  \bibinfo {author} {\bibfnamefont {F.}~\bibnamefont {Wang}},\ and\ \bibinfo
  {author} {\bibfnamefont {X.~C.}\ \bibnamefont {Xie}},\ }\bibfield  {title}
  {\bibinfo {title} {Detection of spinons via spin transport},\ }\href
  {https://doi.org/10.1103/PhysRevB.88.041405} {\bibfield  {journal} {\bibinfo
  {journal} {Phys. Rev. B}\ }\textbf {\bibinfo {volume} {88}},\ \bibinfo
  {pages} {041405} (\bibinfo {year} {2013})}\BibitemShut {NoStop}%
\bibitem [{\citenamefont {Chatterjee}\ and\ \citenamefont
  {Sachdev}(2015)}]{Chatterjee2015}%
  \BibitemOpen
  \bibfield  {author} {\bibinfo {author} {\bibfnamefont {S.}~\bibnamefont
  {Chatterjee}}\ and\ \bibinfo {author} {\bibfnamefont {S.}~\bibnamefont
  {Sachdev}},\ }\bibfield  {title} {\bibinfo {title} {Probing excitations in
  insulators via injection of spin currents},\ }\href
  {https://doi.org/10.1103/PhysRevB.92.165113} {\bibfield  {journal} {\bibinfo
  {journal} {Phys. Rev. B}\ }\textbf {\bibinfo {volume} {92}},\ \bibinfo
  {pages} {165113} (\bibinfo {year} {2015})}\BibitemShut {NoStop}%
\bibitem [{\citenamefont {de~Carvalho}\ \emph {et~al.}(2018)\citenamefont
  {de~Carvalho}, \citenamefont {Freire}, \citenamefont {Miranda},\ and\
  \citenamefont {Pereira}}]{DeCarvalho2018}%
  \BibitemOpen
  \bibfield  {author} {\bibinfo {author} {\bibfnamefont {V.~S.}\ \bibnamefont
  {de~Carvalho}}, \bibinfo {author} {\bibfnamefont {H.}~\bibnamefont {Freire}},
  \bibinfo {author} {\bibfnamefont {E.}~\bibnamefont {Miranda}},\ and\ \bibinfo
  {author} {\bibfnamefont {R.~G.}\ \bibnamefont {Pereira}},\ }\bibfield
  {title} {\bibinfo {title} {{Edge magnetization and spin transport in an
  SU(2)-symmetric Kitaev spin liquid}},\ }\href
  {https://doi.org/10.1103/PhysRevB.98.155105} {\bibfield  {journal} {\bibinfo
  {journal} {Phys. Rev. B}\ }\textbf {\bibinfo {volume} {98}},\ \bibinfo
  {pages} {155105} (\bibinfo {year} {2018})}\BibitemShut {NoStop}%
\end{thebibliography}%



\end{document}

\title{Supplemental Material for Spin Transport in a Quantum Spin-Orbital Liquid}
\author{Zekun Zhuang}
\affiliation{Department of Physics,   Brown University, Providence, Rhode Island 02912-1843, USA}
\author{J. B. Marston}
\affiliation{Department of Physics,   Brown University, Providence, Rhode Island 02912-1843, USA}
\affiliation{Brown Theoretical Physics Center, Brown University, Providence, Rhode Island 02912-1843, USA}
\maketitle

In this supplemental material, we provide detailed calculations that support the results in the main text. In Sec. \ref{NEGF}, we define and derive the Green's functions as well as the self-energies in real-time formalism. We explicitly calculate the first-order self-energies $\Sigma^{f(1)}$ assuming periodic boundary condition along $y$ direction in Sec. \ref{firstorder}. In Sec. \ref{sc} we discuss the effects of the self-consistent self-energies $\Sigma^{f(sc)}$. We discuss the relation of the spin current expressions between this work and previous work in Sec. \ref{relation}.
\section{Non-equilibrium Green's functions}\label{NEGF}
We define the following contour-ordered bare Green's functions
\begin{equation}
   g^f_{ij}(t,t^\prime)=-i\langle \mathcal{T}_c f_{i,z}(t)f_{j,z}^\dagger(t^\prime)\rangle,
\end{equation}
\begin{equation}
   g^{\gamma}_{ij}(t,t^\prime)=-i\langle \mathcal{T}_c \gamma_i^z(t)\gamma_j^z(t^\prime)\rangle,
\end{equation}
\begin{equation}
    g^M_{\alpha,ij}(t,t^\prime)=-i\sum_{i_\alpha,j_\alpha}\langle \mathcal{T}_c S_{i_\alpha }^-(t)S_{j_\alpha}^+(t^\prime)\rangle
\end{equation}
in the interaction picture. In the real-time formalism, the various bare Green's functions are hence
\begin{equation}
    \begin{split}
          &g^{f,R}_{ij}(t,t^\prime)=- i\theta(t-t^\prime)\langle\{f_{i,z}(t),f_{j,z}^\dagger(t^\prime)\}\rangle, \\
                    &g^{f,A}_{ij}(t,t^\prime)= i\theta(t^\prime-t)\langle\{f_{i,z}(t),f_{j,z}^\dagger(t^\prime)\}\rangle, \\
    &g^{f,K}_{ij}(t,t^\prime)=-i\langle[f_{i,z}(t),f_{j,z}^\dagger(t^\prime)]\rangle,  \\
    \end{split}
\end{equation}
\begin{equation}
    \begin{split}
          &g^{\gamma,R}_{ij}(t,t^\prime)=- i\theta(t-t^\prime)\langle\{\gamma_i^z(t),\gamma_j^z(t^\prime)\}\rangle, \\
                    &g^{\gamma,A}_{ij}(t,t^\prime)= i\theta(t^\prime-t)\langle\{\gamma_i^z(t),\gamma_j^z(t^\prime)\}\rangle ,\\
    &g^{\gamma,K}_{ij}(t,t^\prime)=-i\langle[\gamma_i^z(t),\gamma_j^z(t^\prime)]\rangle , \\
    \end{split}
\end{equation}
\begin{equation}
    \begin{split}
          &g^{M,R}_{\alpha,ij}(t,t^\prime)=- i\theta(t-t^\prime)\sum_{i_\alpha,j_\alpha}\langle [S_{i_\alpha}^-(t),S_{j_\alpha}^+(t^\prime)]\rangle, \\
                    &g^{M,A}_{\alpha,ij}(t,t^\prime)= i\theta(t^\prime-t)\sum_{i_\alpha,j_\alpha}\langle[S_{i_\alpha}^-(t),S_{j_\alpha}^+(t^\prime)]\rangle, \\
    &g^{M,K}_{\alpha,ij}(t,t^\prime)=-i\sum_{i_\alpha,j_\alpha}\langle\{S_{i_\alpha}^-(t),S_{j_\alpha}^+(t^\prime)\}\rangle.  \\ 
    \end{split}
    \label{gMRAK}
\end{equation}
In the frequency domain, the retarded/advanced (R/A) Green's functions for $g^f$ is obvious:
\begin{equation}
   \bm{g}^{f,R/A}(\omega)=(\omega-\bm{h}\pm i\delta)^{-1},
\end{equation}
where we use bold symbols to denote the matrices and $\bm{h}$ is defined in the main text. The retarded/advanced Green's functions for $g^\gamma$ may be obtained by making the following transformations
\begin{equation}
    c_\lambda^\dagger=\frac{1}{\sqrt{2}} \sum_i\langle i\vert \lambda\rangle \gamma_i^z,
\end{equation}
\begin{equation}
    \gamma_i^z=\sqrt{2}\sum_\lambda \left(\langle \lambda\vert i\rangle c_\lambda^\dagger+\langle i\vert \lambda\rangle c_{\lambda}\right),
\end{equation} 
so that the Majorana Hamiltonian is diagonalized in terms of complex fermions $c_\lambda$
\begin{equation}
    H_\text{K}=\sum_\lambda \epsilon_\lambda( c_\lambda^\dagger c_\lambda-\frac{1}{2}),
\end{equation}
where $\epsilon_\lambda$ labels the positive eigenvalues of matrix $\bm{h}$ and $\vert \lambda\rangle$ is the corresponding normalized eigenvector. After making such transformations, one can show that
\begin{equation}
    g_{ij}^{\gamma,R/A}(\omega)=2\sum_\lambda\left(\frac{\langle i\vert \lambda\rangle\langle\lambda\vert j\rangle}{\omega-\epsilon_\lambda \pm i\delta }+\frac{\langle j\vert \lambda\rangle\langle\lambda\vert i\rangle}{\omega+\epsilon_\lambda\pm i\delta}\right).
\end{equation}
Note that for skew-symmetric matrix $\bm{h}$ the eigenvectors $\vert \pm \lambda\rangle$ with opposite eigenvalues $\pm \epsilon_\lambda$ are related by $\langle i\vert \lambda\rangle=\langle -\lambda\vert i\rangle e^{i\phi_\lambda}$, therefore the above equation can also be rewritten as 
\begin{equation}
    \bm{g}^{\gamma,R/A}(\omega)=2(\omega-\bm{h}\pm i\delta)^{-1}.
\end{equation}
The Keldysh (K) component may be obtained via the fluctuation-dissipation (FT) theorem 
\begin{equation}
    \bm{g}^{f(\gamma),K}(\omega)=\tanh\frac{\beta \omega}{2}[\bm{g}^{f(\gamma),R}(\omega)-\bm{g}^{f(\gamma),A}(\omega)]. \label{FTtheoremFermi}
\end{equation}
Following the definition Eq. (\ref{gMRAK}) one can show that
\begin{equation}
    \zz{g^{M,R/A}_{\alpha,ij}(\omega)=\sum_{i_\alpha,j_\alpha}\sum_{n,n^\prime}\langle n\vert i_\alpha\rangle\langle j_\alpha\vert n\rangle \langle i_\alpha\vert n^\prime \rangle\langle n^\prime\vert j_\alpha\rangle\frac{n_\downarrow({\epsilon_{n}})-n_\uparrow({\epsilon_{n^\prime}})}{\omega+\epsilon_{n}-\epsilon_{n^\prime}\pm i\delta},} \label{gMRAij}
\end{equation}
and
\begin{equation}
     \zz{g^{M,K}_{\alpha,ij}(\omega)=-2\pi i\sum_{i_\alpha,j_\alpha}\sum_{n,n^\prime}\langle n\vert i_\alpha\rangle\langle j_\alpha\vert n\rangle \langle i_\alpha\vert n^\prime \rangle\langle n^\prime\vert j_\alpha\rangle\left[n_\downarrow({\epsilon_{n}})(1-n_\uparrow({\epsilon_{n^\prime}}))+n_\uparrow({\epsilon_{n^\prime}})(1-n_\downarrow({\epsilon_{n}})) \right]\delta(\omega+\epsilon_{n}-\epsilon_{n^\prime}),}
\end{equation}
where \zz{$n,n^\prime$} label spin bath's eigenstate and \zz{$n_{\uparrow(\downarrow)}(...)$ denotes the occupancy for spin up (down) states}. One can check that
\begin{equation}
        \bm{g}^{M,K}_\alpha(\omega)=\coth\frac{\beta (\omega-V_\alpha)}{2}[\bm{g}^{M,R}_\alpha(\omega)-\bm{g}^{M,A}_\alpha(\omega)].\label{FTtheoremBose}
\end{equation}
The full Green's functions on the Keldysh contour are defined as
\begin{equation}
   G^f_{ij}(t,t^\prime)=-i\langle \mathcal{T}_c e^{-i\int_c d\tau H_{\text{int}}(\tau)} f_{i,z}(t)f_{j,z}^\dagger(t^\prime)\rangle,
\end{equation}
\begin{equation}
   G^{\gamma}_{ij}(t,t^\prime)=-i\langle \mathcal{T}_c e^{-i\int_c d\tau H_{\text{int}}(\tau)} \gamma_i^z(t)\gamma_j^z(t^\prime)\rangle.
\end{equation}
Note that the full Green's functions also only depend on the time difference $t-t^\prime$ if the system is in the NESS. In the real-time formalism, their R/A/K components are given by
\begin{equation}
   \bm{G}^{f,R/A}(\omega)=\left(\omega-\bm{h}-\bm{\Sigma}^{f,R/A}(\omega)\right)^{-1}, \label{GfRA}
\end{equation}
\begin{equation}
   \bm{G}^{\gamma,R/A}(\omega)=\left(\frac{\omega-\bm{h}}{2}-\bm{\Sigma}^{\gamma,R/A}(\omega)\right)^{-1}, \label{GgammaRA}
\end{equation}
\begin{equation}
   \bm{G}^{f(\gamma),K}(\omega)=\bm{G}^{f(\gamma),R}(\omega)\bm{\Sigma}^{f(\gamma),K}(\omega)\bm{G}^{f(\gamma),A}(\omega). \label{KeldyshE}
\end{equation}
For the diagrams in the main text, the R/A/K components of the self-energies $\Sigma^{f(1)}$ can be obtained via the \textit{RAK-rules} \cite{rammer2007}:
\begin{equation}
        \Sigma^{f(1),R/A}_{\alpha,ij}(\omega)=\frac{i(\lambda^\alpha)^2}{4}\int \frac{d\omega^\prime}{2\pi}\frac{1}{2}\left[g_{\alpha,ij}^{M,R/A}(\omega^\prime)g_{\alpha,ij}^{\gamma,K}(\omega-\omega^\prime)+g_{\alpha,ij}^{M,K}(\omega^\prime)g_{\alpha,ij}^{\gamma,R/A}(\omega-\omega^\prime)\right], \label{Sigf1RA}
\end{equation}
\begin{equation}
    \Sigma^{f(1),K}_{\alpha,ij}(\omega)=\frac{i(\lambda^\alpha)^2}{4}\int \frac{d\omega^\prime}{2\pi}\frac{1}{2}
\left[\left(g_{\alpha,ij}^{M,R}(\omega^\prime)-g_{\alpha,ij}^{M,A}(\omega^\prime)\right)\left(g_{\alpha,ij}^{\gamma,R}(\omega-\omega^\prime)-g_{\alpha,ij}^{\gamma,A}(\omega-\omega^\prime)\right)\right.
\\+\left.g_{\alpha,ij}^{M,K}(\omega^\prime)g_{\alpha,ij}^{\gamma,K}(\omega-\omega^\prime)\right].\label{Sigf1K}
\end{equation}
Similarly the self-energy $\Sigma^{f(sc)}$ can be obtained by (\ref{Sigf1RA}) and (\ref{Sigf1K}), with $g^\gamma$ being replaced by $G^\gamma$. The self-energy $\Sigma^{\gamma(sc)}$ is given by

\begin{equation}
\begin{split}
     \Sigma^{\gamma(sc),R/A}_{\alpha,ij}(\omega)=\frac{i(\lambda^\alpha)^2}{4}\int \frac{d\omega^\prime}{2\pi}\frac{1}{2}&\left[g^{M,A/R}_{\alpha,ji}(\omega^\prime)G^{f,K}_{ij}(\omega+\omega^\prime)+g^{M,K}_{\alpha,ji}(\omega^\prime)G^{f,R/A}_{ij}(\omega+\omega^\prime)\right.\\&\left.-g^{M,R/A}_{\alpha,ij}(\omega^\prime)G^{f,K}_{ji}(\omega^\prime-\omega)-g^{M,K}_{\alpha,ij}(\omega^\prime)G^{f,A/R}_{ji}(\omega^\prime-\omega)\right],  \\ 
\end{split}
\label{SignuRA}
\end{equation}

\begin{widetext}
\begin{equation}
\begin{split}
     \Sigma^{\gamma(sc),K}_{\alpha,ij}(\omega)=\frac{i(\lambda^\alpha)^2}{4}\int \frac{d\omega^\prime}{2\pi}\frac{1}{2}&\left[\left(g^{M,A}_{\alpha,ji}(\omega^\prime)-g^{M,R}_{\alpha,ji}(\omega^\prime)\right)\left(G^{f,R}_{ij}(\omega+\omega^\prime)-G^{f,A}_{ij}(\omega+\omega^\prime)\right)\right.\\&-\left(g^{M,R}_{\alpha,ij}(\omega^\prime)-g^{M,A}_{\alpha,ij}(\omega^\prime)\right)\left(G^{f,A}_{ji}(\omega^\prime-\omega)-G^{f,R}_{ji}(\omega^\prime-\omega)\right)  \\ 
     &+\left.g^{M,K}_{\alpha,ji}(\omega^\prime)G^{f,K}_{ij}(\omega+\omega^\prime)-g^{M,K}_{\alpha,ij}(\omega^\prime)G^{f,K}_{ji}(\omega^\prime-\omega)\right].\\
\end{split}
\label{SignuK}
\end{equation}
\end{widetext}
According to the definitions, the (full) Green's functions have the following properties:
\begin{equation}
    \bm{G}^{f(\gamma,M),R}(\omega)=\left[\bm{G}^{f(\gamma,M),A}(\omega)\right]^\dagger,
\end{equation}
\begin{equation}
    \bm{G}^{f(\gamma,M),K}(\omega)=-\left[\bm{G}^{f(\gamma,M),K}(\omega)\right]^\dagger,
\end{equation}
\begin{equation}
    G^{\gamma,R/A}_{ij}(\omega)=-\left[G^{\gamma,R/A}_{ij}(-\omega)\right]^*, \label{MajoranaPHrelation1}
\end{equation}
\begin{equation}
      G^{\gamma,K}_{ij}(\omega)=\left[G^{\gamma,K}_{ij}(-\omega)\right]^*. \label{MajoranaPHrelation2}  
\end{equation}
Note that Eq. (\ref{MajoranaPHrelation1}) and Eq. (\ref{MajoranaPHrelation2}) arises from the intrinsic particle-hole (PH) symmetry of Majorana fermions, which is also protected by the PH symmetry of the self-energies (note $\Sigma^{\gamma,R/A}_{ij}(\omega)=-\left[\Sigma^{\gamma,R/A}_{ij}(-\omega)\right]^*$ and $\Sigma^{\gamma,K}_{ij}(\omega)=\left[\Sigma^{\gamma,K}_{ij}(-\omega)\right]^*$ from Eq. (\ref{SignuRA}) and Eq. (\ref{SignuK})). 
 
\section{spectral function and distribution function of the FM baths calculated by \texorpdfstring{$\Sigma^{f(1)}$}{} }\label{firstorder}
We consider the zigzag-type contact (ZC), as shown in Figure 4 in the main text. We impose periodic boundary condition along the $y$ direction and assume that along this direction the system plus bath are translationally invariant. In this context one can show that Eq. (\ref{gMRAij}) becomes
\begin{equation}
\begin{split}
    g^{M,R/A}_{\alpha}(k,\omega)
    &=\frac{1}{N_y}\sum_q\int d\epsilon_q d\epsilon_{k+q} J^\alpha_\downarrow(q,\epsilon_q)J_\uparrow^\alpha(k+q,\epsilon_{k+q})\zz{\frac{n_\downarrow({\epsilon_{q}})-n_\uparrow({\epsilon_{k+q }})}{\omega+\epsilon_{q}-\epsilon_{k+q }\pm i\delta}}\\    
    &\approx\frac{1}{N_y}\sum_q J^\alpha_\downarrow(q)J_\uparrow^\alpha(k+q)\int d\omega^\prime\frac{\omega^\prime-V_\alpha}{\omega-\omega^\prime\pm i\delta},\\
\end{split}
\label{gMRAky}
\end{equation}
where $\epsilon_q$ $(\vert \epsilon_q \rangle)$ is the eigenenergy (eigenstate) of the spin bath at transverse momentum $q$, $J_{\uparrow(\downarrow)}^\alpha(q,\omega)$ is the local density of states (LDOS) of up (down) spin band at the boundary of the spin bath $\alpha$ in momentum and frequency space. Note that in the second line we have assumed $V_\alpha$ and $T$ are much smaller than the variation scale of $J^\alpha$ and ignore the $\omega$-dependence of $J^\alpha(q,\omega)$.
With Eqs. (\ref{FTtheoremFermi})(\ref{FTtheoremBose})(\ref{Sigf1RA}) and (\ref{gMRAky}) one can show that the imaginary part of the self-energy $\text{Im}\Sigma^{f(1),R}_{\alpha,ii}(k_y,\omega)$ ($i\in A_\alpha$) at low energies is given by

\begin{equation}
\begin{split}
    \text{Im}\Sigma^{f(1),R}_{\alpha,ii}(k_y,\omega)&=\frac{(\lambda^\alpha)^2}{8N_y^2} \sum_{k^\prime,q}J_\uparrow^\alpha(k^\prime+q) J_\downarrow^\alpha(q)\int d\omega^\prime  (\omega^\prime-V_\alpha)\\ &\times\left[\tanh \frac{\beta(\omega-\omega^\prime)}{2}+\coth \frac{\beta(\omega^\prime-V_\alpha)}{2}\right]\text{Im} g^{\gamma,R}_{\alpha,ii}(k_y-k^\prime,\omega-\omega^\prime).\\
\end{split}
\label{SigfRexpression}
\end{equation}

 The real part of the self-energy can be obtained by Kramers-Kronig relations \mbox{$\text{Re}\Sigma^{f(1),R}_\alpha(k_y,\omega)=-\frac{1}{\pi}\mathcal{P}\int d\omega^\prime\frac{\text{Im}\Sigma^{f(1),R}_\alpha(k_y,\omega^\prime)}{\omega-\omega^\prime}$} and it depends on the details of spin bath. The Keldysh self-energy can be obtained by combining Eqs. (\ref{FTtheoremFermi})(\ref{FTtheoremBose})(\ref{Sigf1K}) and (\ref{gMRAky})

\begin{equation}
\begin{split}
    \text{Im}\Sigma^{f(1),K}_{\alpha,ii}(k_y,\omega)&=\frac{(\lambda^\alpha)^2}{4N_y^2} \sum_{k^\prime,q}J_\uparrow^\alpha(k^\prime+q) J_\downarrow^\alpha(q)\int d\omega^\prime   (\omega^\prime-V_\alpha)\\
    &\times\left[1+\tanh \frac{\beta(\omega-\omega^\prime)}{2}\coth \frac{\beta(\omega^\prime-V_\alpha)}{2}\right]\text{Im} g^{\gamma,R}_{\alpha,ii}(k_y-k^\prime,\omega-\omega^\prime).\\ 
\end{split}
\label{SigfKexpression}
\end{equation}

Notice that $(1+\tanh x\coth y)/(\tanh x+\coth y)=\tanh (x+y)$, combine Eq. (\ref{SigfRexpression}) and Eq. (\ref{SigfKexpression}), one obtains $f^{\alpha(1)}_{\text{eff}}(\omega)=\frac{1}{e^{\beta(\omega-V_\alpha)}+1}$. 
Within the LSEA, or assuming that $J^\alpha_{\uparrow(\downarrow)}=J^\alpha$ is $k$-independent, the local self-energy is given by
\begin{widetext}
\begin{equation}
    \text{Im}\Sigma^{f(1),R}_{\alpha,ii}(\omega)
   =-\frac{\pi(\lambda^\alpha J^\alpha)^2}{4} \int d\omega^\prime  \left[\tanh \frac{\beta(\omega-\omega^\prime)}{2}+\coth \frac{\beta(\omega^\prime-V_\alpha)}{2}\right] (\omega^\prime-V_\alpha)D^\alpha_{ii}(\omega-\omega^\prime).
\end{equation}
\end{widetext}
For an armchair-type contact (AC), the form of self-energy is in general more complicated and we will not derive it explicitly. Within the LSEA, however it is given by a similar formula as that in the zigzag case. We remark that such approximation should not affect our results qualitatively as long as it does not change the $\omega$-dependence of $\text{Im}\Sigma^{f,R}(\omega)$ at low-energies, and hence does not change, for example the power law of $I_s-V_s$ relation.

\section{Effects of \texorpdfstring{$\Sigma^{f(sc)}$}{}}\label{sc}

For $\Sigma^{f(sc)}$, one needs to combine Eqs. (\ref{GfRA})-(\ref{SignuK}) and do self-consistent calculations. We first consider the case that the system is only in contact with a single spin bath that has spin bias $V_\alpha$ and temperature $T$. In this case, the Majorana sector would be at thermal equilibrium with temperature $T$, while the FM sector is at the equilibrium state with temperature $T$ and chemical potential $V_\alpha$. This can be checked by assuming the above statements are true, plugging the FT theorem
\begin{equation}
        \bm{G}^{f,K}(\omega)=\tanh\frac{\beta (\omega-V_\alpha)}{2}[\bm{G}^{f,R}(\omega)-\bm{G}^{f,A}(\omega)], \label{GfFTtheorem}
\end{equation}
\begin{equation}
        \bm{G}^{\gamma,K}(\omega)=\tanh\frac{\beta \omega}{2}[\bm{G}^{\gamma,R}(\omega)-\bm{G}^{\gamma,A}(\omega)], \label{GgammaFTtheorem}
\end{equation}
and Eq. (\ref{FTtheoremBose}) to Eqs. (\ref{Sigf1RA})-(\ref{SignuK}). Then one would find
\begin{equation}
            \bm{\Sigma}^{f,K}(\omega)=\tanh\frac{\beta (\omega-V_\alpha)}{2}[\bm{\Sigma}^{f,R}(\omega)-\bm{\Sigma}^{f,A}(\omega)],
\end{equation}
\begin{equation}
            \bm{\Sigma}^{\gamma,K}(\omega)=\tanh\frac{\beta \omega}{2}[\bm{\Sigma}^{\gamma,R}(\omega)-\bm{\Sigma}^{\gamma,A}(\omega)],
\end{equation}
which in turn make Eq. (\ref{GfFTtheorem}) and Eq. (\ref{GgammaFTtheorem}) hold.

\begin{figure}[t]
  \subfigure[]{
    \label{fig:ACspec} %% label for second subfigure
    \includegraphics[width=0.45 \textwidth]{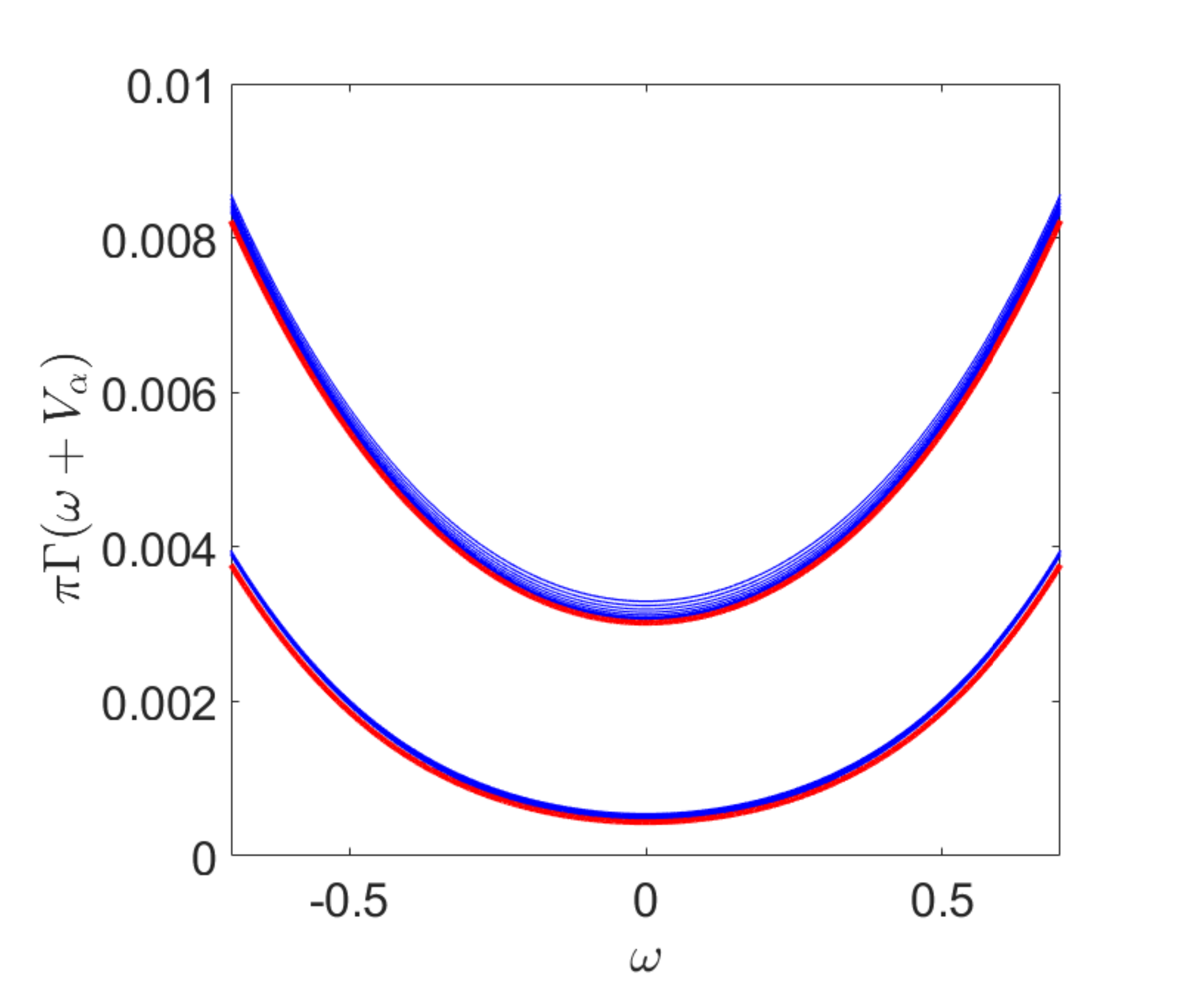}}
          \subfigure[]{
    \label{fig:ZCspec} %% label for first subfigure
    \includegraphics[width=0.45 \textwidth]{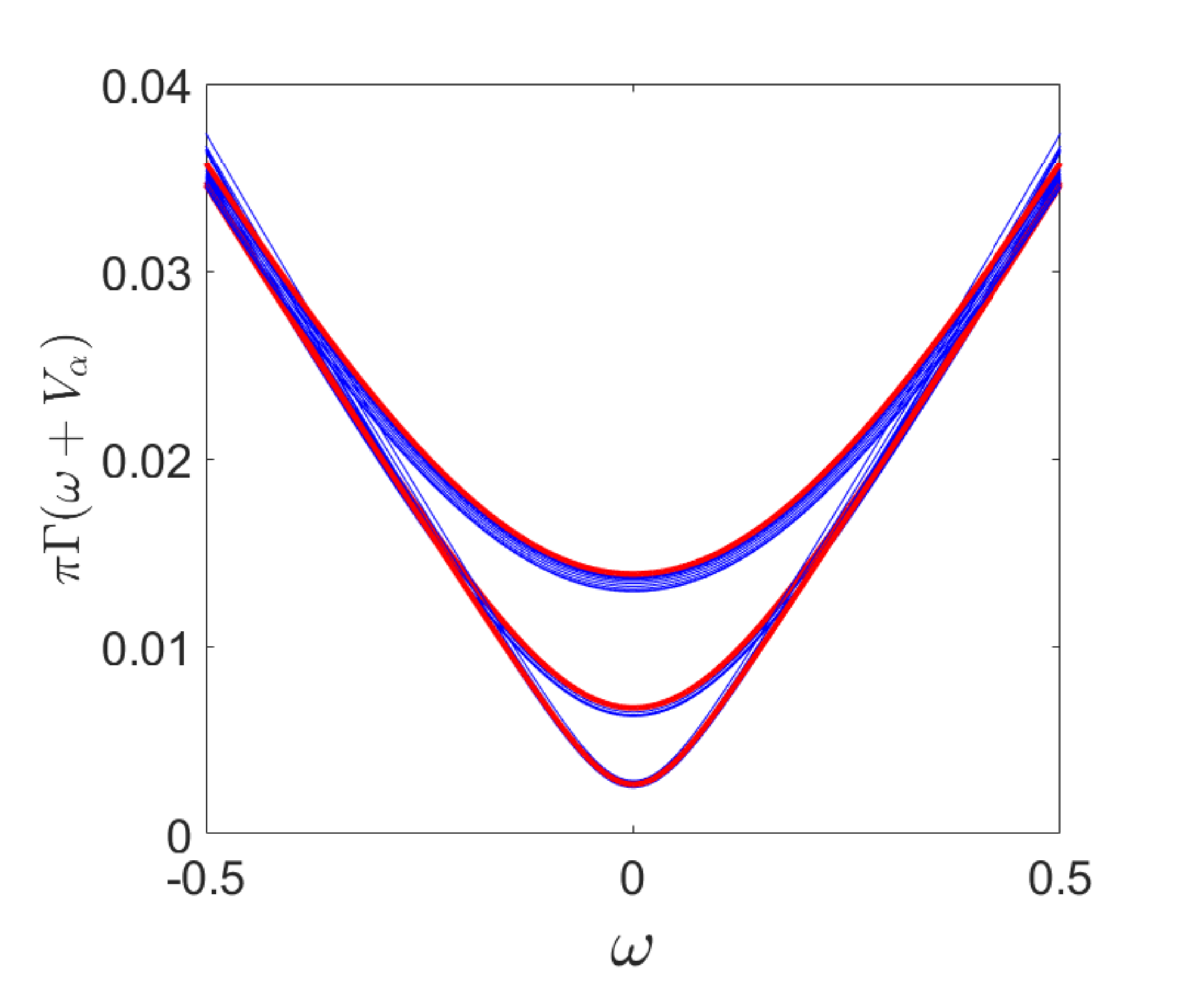}}
    \caption{The spectral function $\Gamma(\omega+V_\alpha)$ of (a) armchair-type contact and (b) zigzag-type contact under LSEA. Each band of blue lines are calculated with $\Sigma^{f(sc)}$ at same temperature $T$ but different spin bias $V_s$ and bath index $\alpha$. The red line close to each band of blue lines are computed with $\Sigma^{f(1)}$ at the same temperature $T$. From bottom to top, the temperatures are (a) $T=0.1, 0.2$, (b) $T=0.02, 0.05, 0.1$. Common parameters: $J_x=J_y=J_z=1$, $\chi=0$, $N_x=20$, $N_y=15$, $(\lambda^\alpha J^\alpha)^2/4=0.1$, $0\leqslant V_s \leqslant 0.4$ }
\label{fig:spec}
\end{figure}

\begin{figure}[t]
  \subfigure[]{
    \label{fig:discomp} %% label for second subfigure
    \includegraphics[height=0.4 \textwidth]{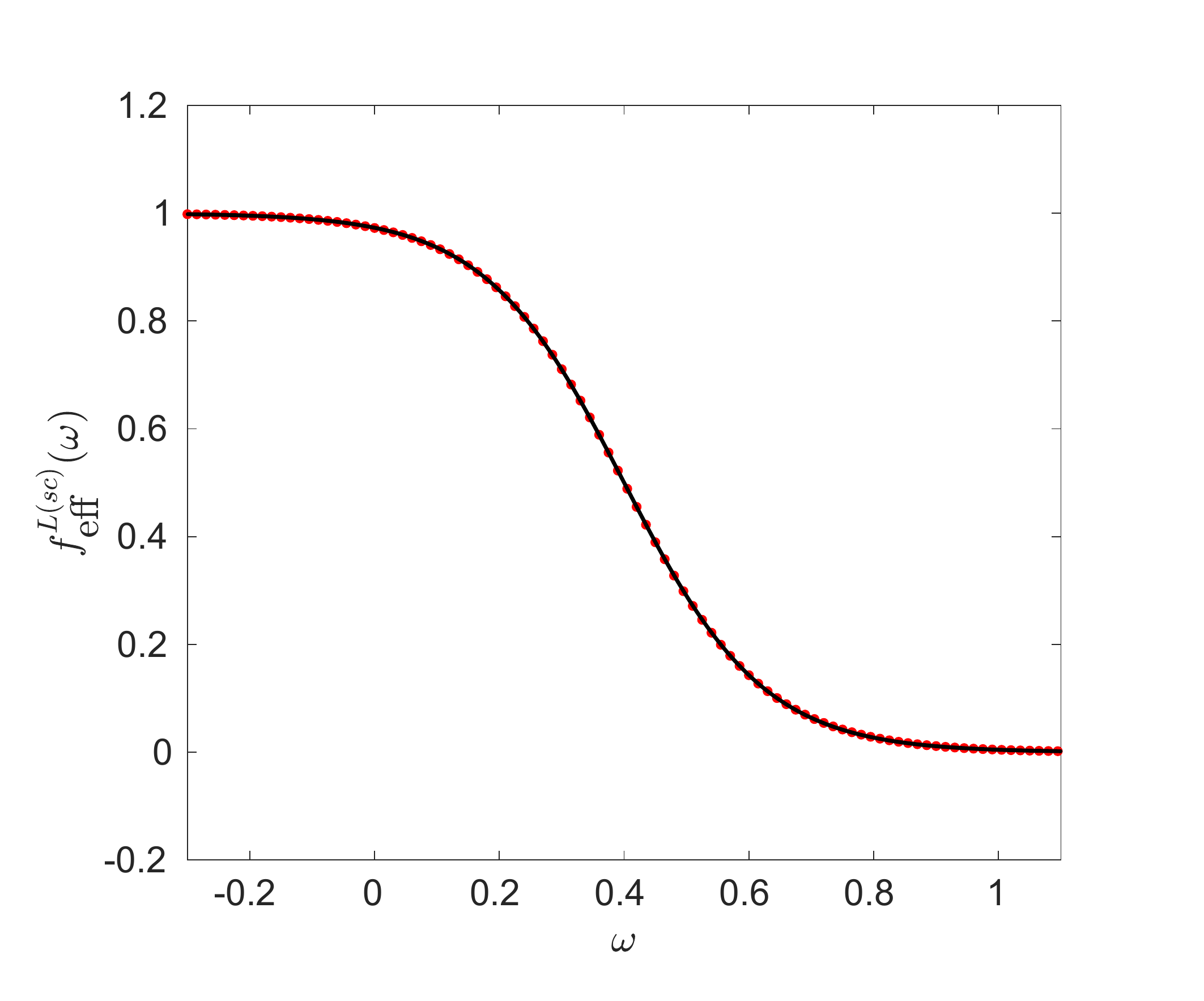}}
          \subfigure[]{
    \label{fig:TV} %% label for first subfigure
    \includegraphics[height=0.4 \textwidth]{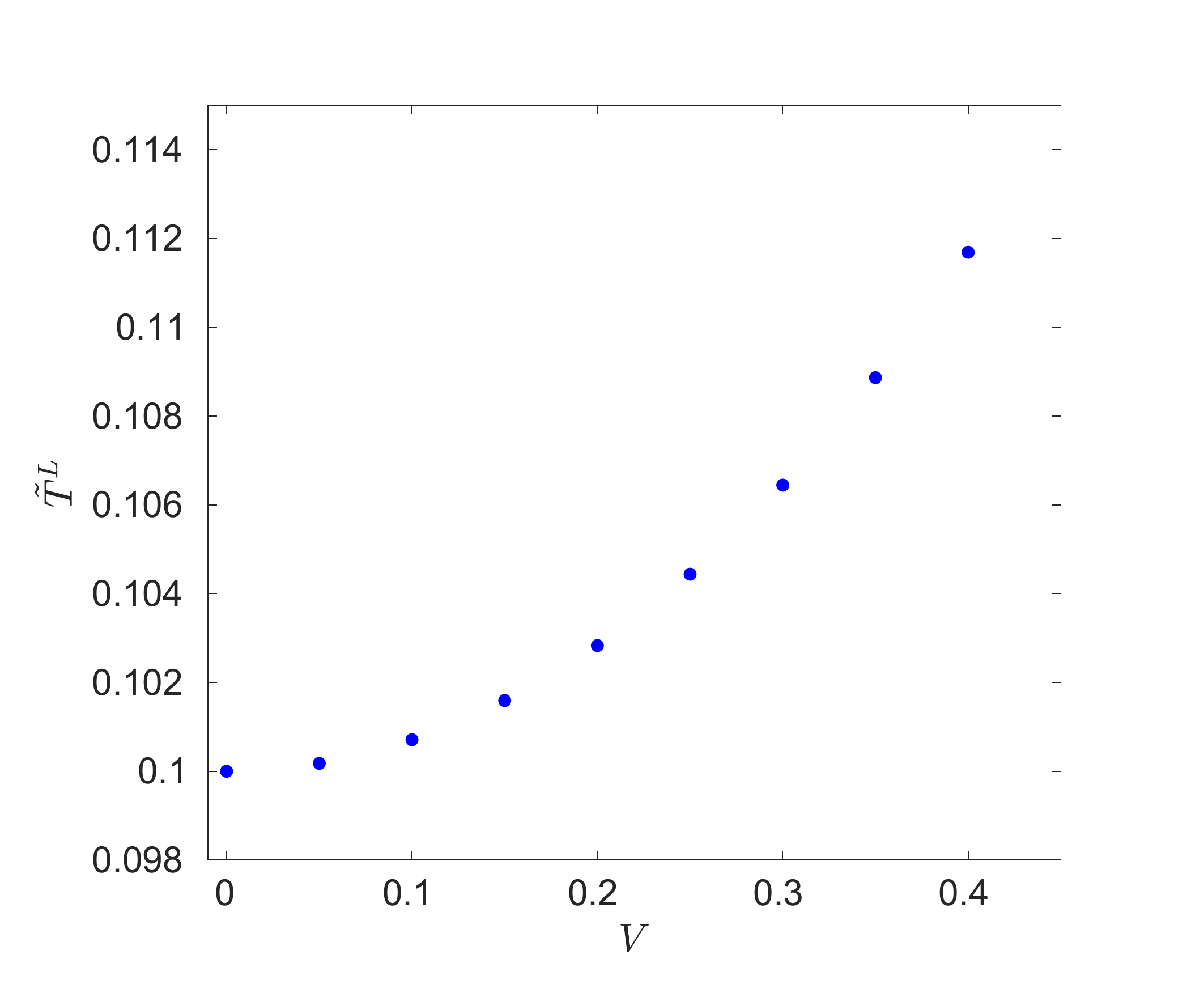}}
    \caption{(a) The effective distribution function $f_\text{eff}^{L(sc)}(\omega)$ of left FM bath (red dots) calculated with $\Sigma^{f(sc)}$ when $V_s=0.4$ and $T=0.1$. The black line is the fit with $\tilde{T}^L\approx 0.1117$. (b) The effective temperature $\tilde{T}^L$ v.s. spin bias $V_s$ when $T=0.1$ (we choose not to present $\tilde{T}^R$ as they are very close). In both figures the contact is armchair-type and LSEA has been used. Common parameters: $J_x=J_y=J_z=1$, $\chi=0$, $N_x=20$, $N_y=15$, $(\lambda^\alpha J^\alpha)^2/4=0.1$. }
\label{fig:dis}
\end{figure}

When the system is connected with multiple spin baths with different spin biases, the problem becomes more complicated. In general, the effective FM bath may not even have a single well-defined distribution function. However when LSEA is used, it can be shown that the effects of $\Sigma^{f(sc)}$ may be just to slightly change the spectral function $\Gamma(\omega)$ (see Fig. \ref{fig:ACspec} and Fig. \ref{fig:ZCspec}) and modify the temperature of the effective FM bath while leave its chemical potential $\mu_\alpha=V_\alpha$ invariant. Under LSEA, it can be shown that at low energies $\text{Im}\Sigma^{f(sc),R}_\alpha(\omega)$ is an even function about $\omega=V_\alpha$ while $\text{Im}\Sigma^{f(sc),K}_\alpha(\omega)$ is odd regarding $\omega=V_\alpha$ as $\text{Im} G^{\gamma,R}(\omega)=\text{Im} G^{\gamma,R}(-\omega)$, which is related to the intrinsic PH symmetry of Majorana fermions. Therefore $\text{Im}\Sigma^{f(sc),K}_\alpha(\omega)/\text{Im}\Sigma^{f(sc),R}_\alpha(\omega)$ is an odd function about $\omega=V_\alpha$ and $f^\alpha(\omega)$ may be approximately expressed as a Fermi distribution function with $\mu_\alpha=V_\alpha$ and $\tilde{T_\alpha}$, as verified in Fig. \ref{fig:discomp} and Fig. \ref{fig:TV}. 
 
\section{Relation with previous work}\label{relation}
In this section we illustrate the relation between our work and previous work in Ref. \cite{Chen2013,Chatterjee2015,DeCarvalho2018}. Particularly we follow the line sketched in Ref. \cite{Chen2013}. The spin current at the left boundary is defined as
\begin{equation}
    I_s=\sum_{\langle i,i_L\rangle\in A_L}\frac{i\lambda^L}{2}\left(S^-_{i}S_{i_L}^+-S_{i_L}^-S^+_{i}\right).
\end{equation}
To the first order of $\lambda^L$, the expectation value of $I_s$ is given by
\begin{equation}
    \langle I_s(t) \rangle=\sum_{i,j\in A_L}\frac{(\lambda^L)^2}{4}\left[g_{\alpha,ji}^M(t,t^\prime)\chi_{ij}(t^\prime,t)-\chi_{ji}(t,t^\prime)g^M_{\alpha,i j}(t^\prime,t)\right]^<
\end{equation}
where we define $\chi_{ij}(t,t^\prime)=-i\langle \mathcal{T}_c S_{i }^-(t)S_{j}^+(t^\prime)\rangle$ and superscript $<(>)$ denotes the lesser (greater) component. With the analytic continuation procedure, one can show that
\begin{equation}
    \langle I_s\rangle=\sum_{i,j\in A_L}\frac{(\lambda^L)^2}{4}\int \frac{d\omega}{2\pi}\left[\chi_{ij}^<(\omega)g^{M,>}_{\alpha,j i}(\omega)-\chi_{ij}^>(\omega)g^{M,<}_{\alpha,j i}(\omega)\right]. \label{spincurrent}
\end{equation}
Note that here $\chi_{ij}^<(\omega)=-i\int dt e^{i\omega t}\langle S_j^+(0)S_i^-(t)\rangle$ and $\chi_{ij}^>(\omega)=-i\int dt e^{i\omega t}\langle S_i^-(t)S_j^+(0)\rangle$ are system's spin correlation functions at equilibrium. Equation (\ref{spincurrent}) is in fact (or equivalent to) the formula that previous work rely on. Further expressing $\chi$ in terms of $g^\gamma$ and $g^f$ one obtains
\begin{equation}
\begin{split}
        \langle I_s\rangle&=\sum_{i,j\in A_L}\frac{i(\lambda^L)^2}{4}\int \frac{d\omega}{2\pi}\frac{d\omega^\prime}{2\pi}\left[g_{ij}^{f,<}(\omega^\prime)g_{\alpha,ji}^{M,>}(\omega)g^{\gamma,<}_{ij}(\omega-\omega^\prime)-g_{ij}^{f,>}(\omega^\prime)g_{\alpha,ji}^{M,<}(\omega)g^{\gamma,>}_{ij}(\omega-\omega^\prime)\right]\\
        &=\sum_{i,j\in A_L}\frac{i(\lambda^L)^2}{4}\int \frac{d\omega}{2\pi}\frac{d\omega^\prime}{2\pi}\left[g_{ij}^{f,>}(\omega^\prime)g_{\alpha,ji}^{M,<}(\omega)g^{\gamma,<}_{ji}(\omega^\prime-\omega)-g_{ij}^{f,<}(\omega^\prime)g_{\alpha,ji}^{M,>}(\omega)g^{\gamma,>}_{ji}(\omega^\prime-\omega)\right]\\
        &=\sum_{i,j\in A_L}\int \frac{d\omega^\prime}{2\pi}\left[g_{ij}^{f,>}(\omega^\prime)\Sigma^{f(1),<}_{L,ji}(\omega^\prime)-g_{ij}^{f,<}(\omega^\prime)\Sigma^{f(1),>}_{L,ji}(\omega^\prime)\right],\\
\end{split}
\end{equation}
where we have used $g^{\gamma,<}_{ij}(\omega)=-g^{\gamma,>}_{ji}(-\omega)$.
It is convenient to assume the contact is zigzag-type and work in $k_y$ space. The above equation then becomes
\begin{equation}
\begin{split}
        \langle I_s\rangle&=\sum_{k_y}\int \frac{d\omega^\prime}{2\pi}\left[g^{f,>}(k_y,\omega^\prime)\Sigma^{f(1),<}_{L}(k_y,\omega^\prime)-g^{f,<}(k_y,\omega^\prime)\Sigma^{f(1),>}_{L}(k_y,\omega^\prime)\right]\\
        &=\sum_{k_y}2\pi\int d\omega \Gamma(k_y,\omega)N(k_y,\omega) \left[f^{L(1)}_\text{eff}(\omega)\left(1-f_\text{eq}(\omega)\right)-f_\text{eq}(\omega)\left(1-f^{L(1)}_\text{eff}(\omega)\right)\right],\\
\end{split}
\end{equation}
where $f_\text{eq}(\omega)=1/(e^{\beta \omega}+1)$ is the distribution function of FMs at equilibrium. The equation is equivalent to one for the tunneling current between the left FM bath and the FM `graphene' at equilibrium, as stated in the main text.

\zz{
\section{Numerical results for $I_s-V_s$ relation}
Fig. \ref{fig:scaling} shows numerical results for the $I_s-V_s$ relation in the gapless phase calculated using the first-order self-energy $\Sigma^{f(1)}$. We verify the expected power laws in different limits $V_s\gg T$ and $V_s\ll T$ for the different types of contact. 
\begin{figure}[!h]
    \subfigure[]{
    \label{subfig:scalinga1}
    \includegraphics[width=0.48\textwidth]{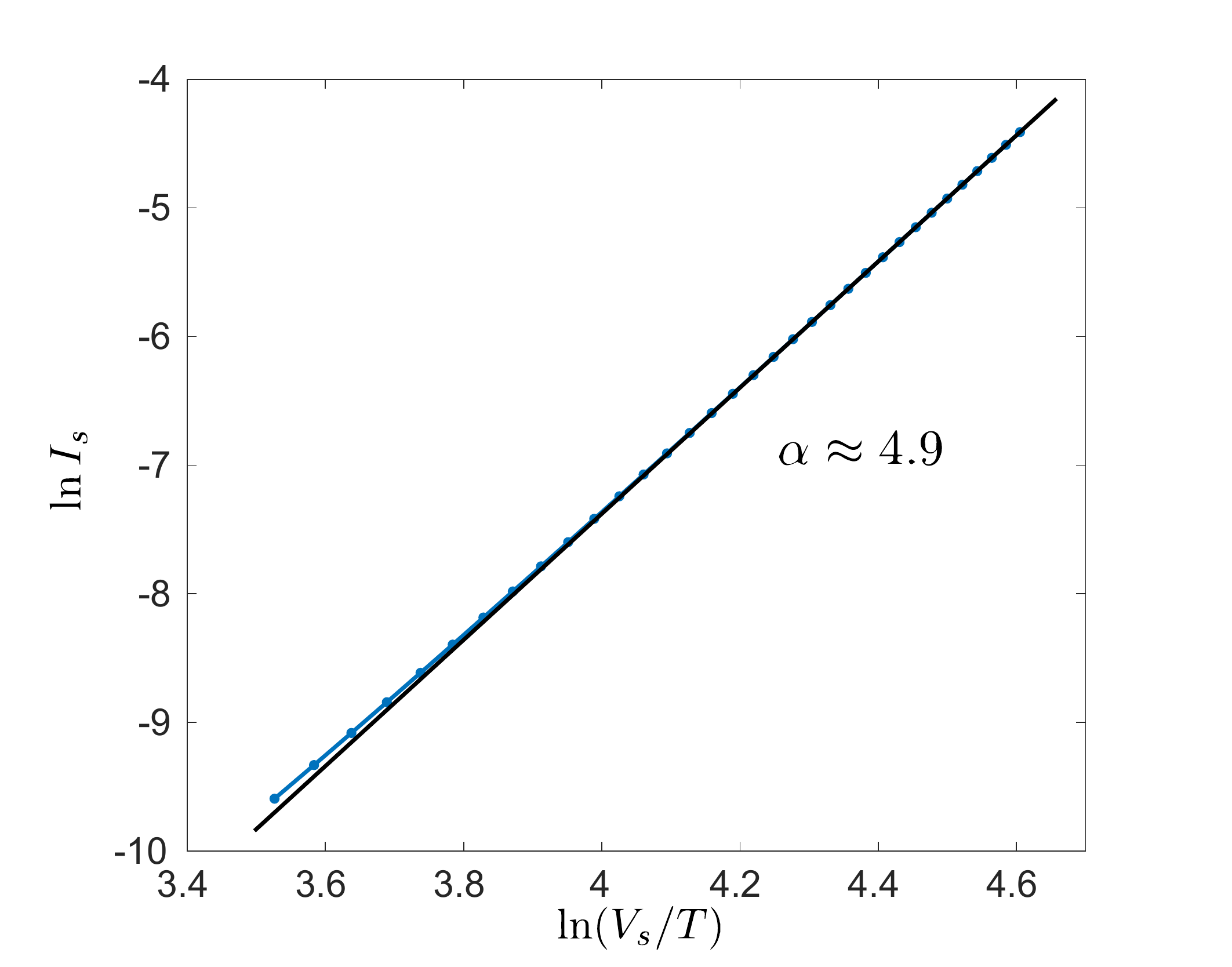}
    }    
    \subfigure[]{
    \label{subfig:scalingz1}
    \includegraphics[width=0.48\textwidth]{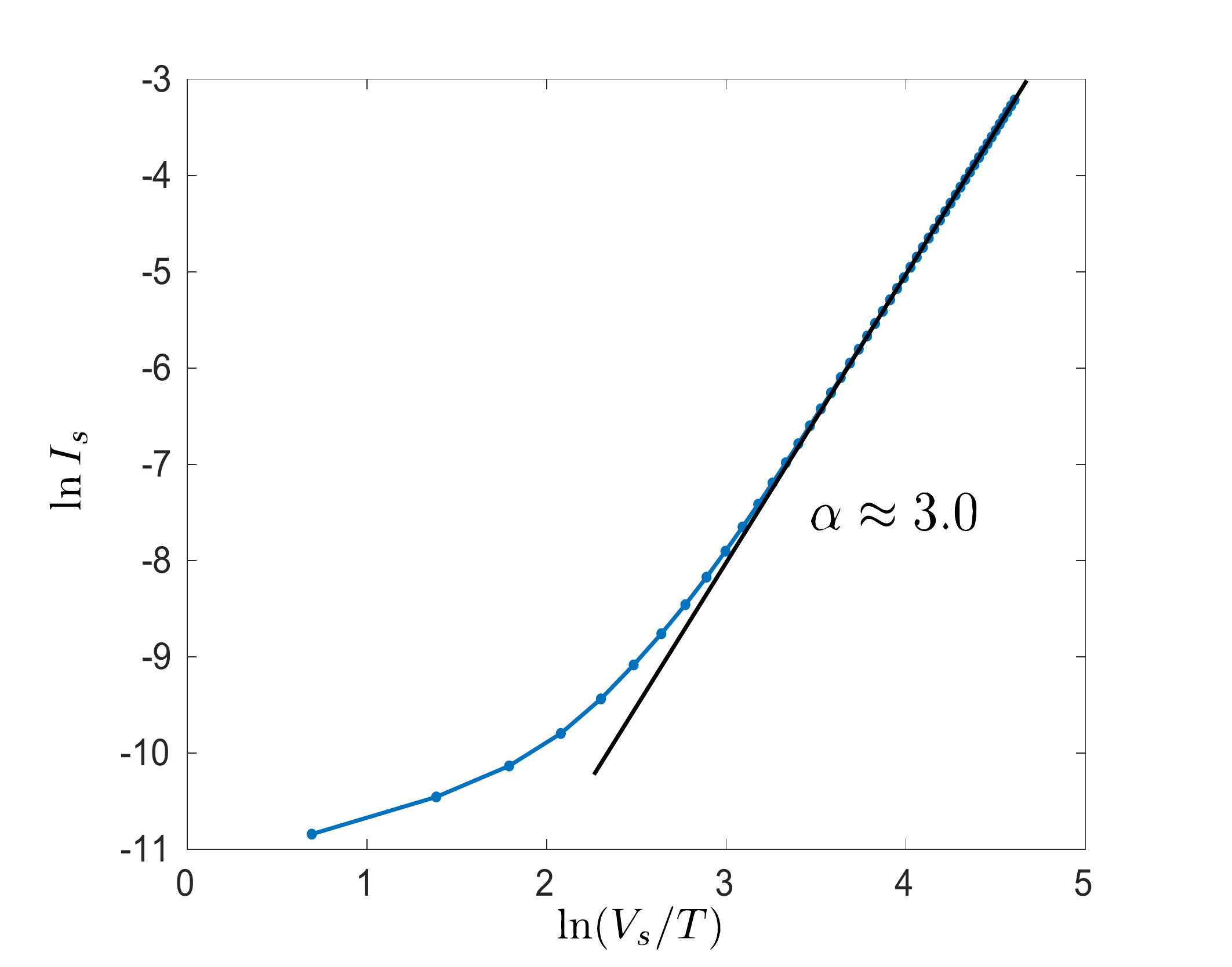}
    }
    \subfigure[]{
    \label{subfig:scalinga2}
    \includegraphics[width=0.48\textwidth]{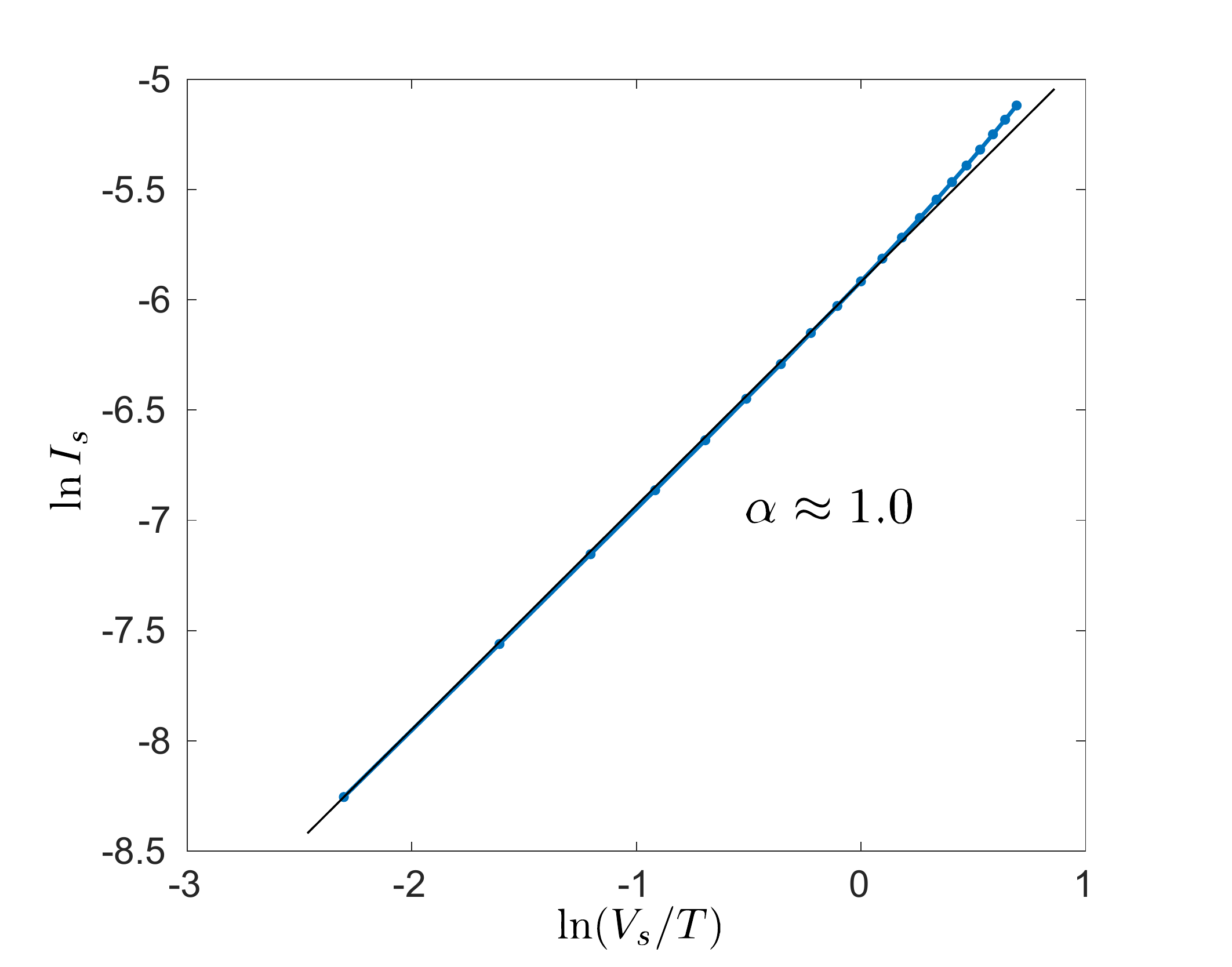}
    }    
    \subfigure[]{
    \label{subfig:scalingz2}
    \includegraphics[width=0.48\textwidth]{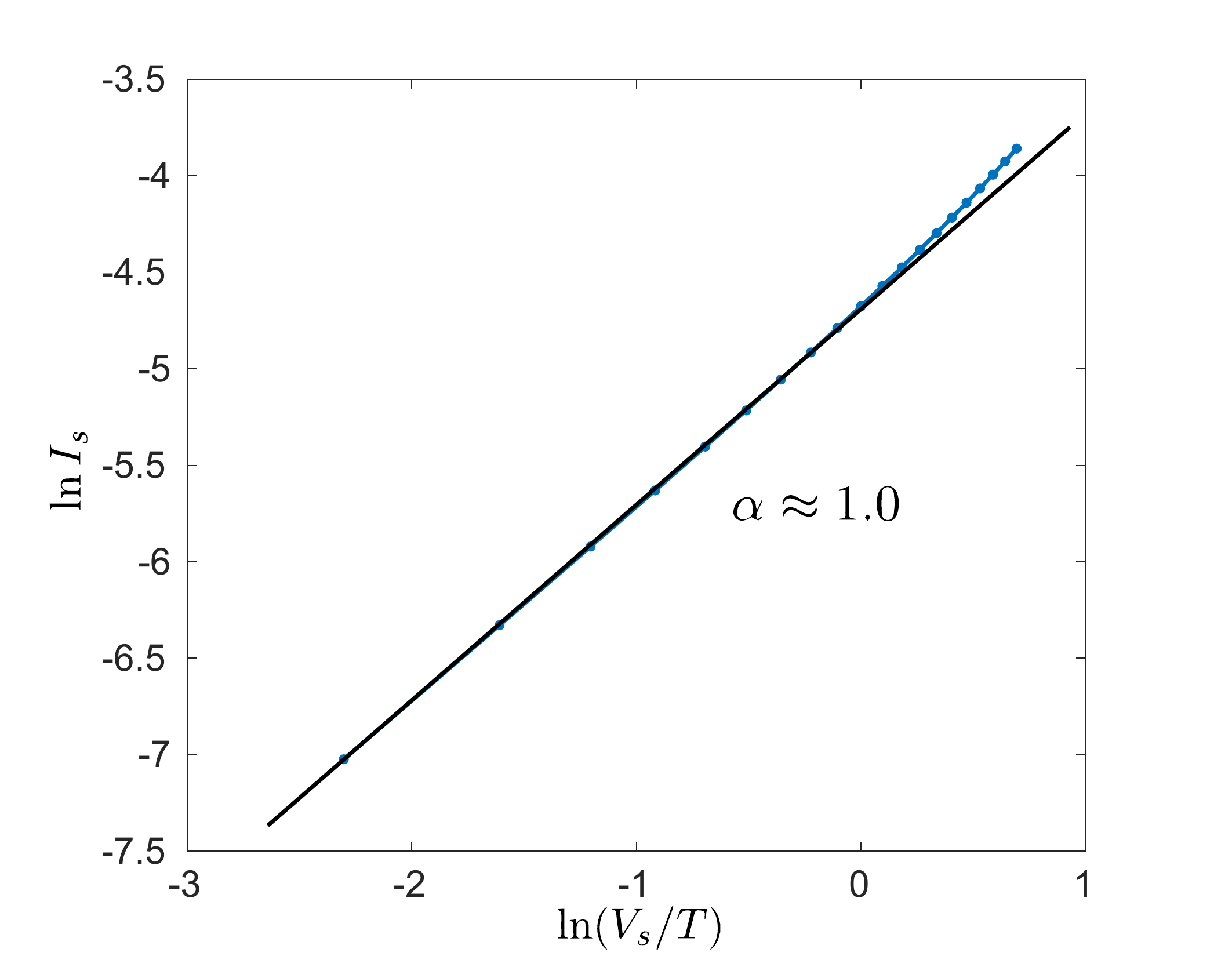}
    }
    \caption{\zz{The numerically calculated $I_s-V_s$ relation in the gapless phase for cases: (a) armchair contact and $V_s\gg T$; (b) zigzag contact and $V_s \gg T$; (c) armchair contact and $V_s\ll T$; and (d) zigzag contact and $V_s \ll T$. The blue dotted lines represent the numerical results while the black solid lines are the linear fits and the fitted slopes $\alpha$ are indicated in the figures. Deviations from predicted power laws at large (small) $V_s$ are due to the transition from the $V_s\ll T$ ($V_s\gg T$) regime to the $V_s\gg T$ ($V_s\ll T$) regime. Parameters common to all the plots are $J_x=J_y=J_z=1$, $N_x=200$. Plot specific parameters are: (a) $N_y=100$, $T=0.01$, $(\lambda^\alpha J^\alpha)^2/4=1$; (b) $N_y=150$, $T=0.01$, $(\lambda^\alpha J^\alpha)^2/4=0.1$; (c) $N_y=100$, $T=0.2$, $(\lambda^\alpha J^\alpha)^2/4=0.1$; and (d) $N_y=150$, $T=0.2$, $(\lambda^\alpha J^\alpha)^2/4=0.1$.}}
    \label{fig:scaling}
\end{figure}
} 
\bibliography{supplementalBib}